\documentclass[fleqn,usenatbib,onecolumn]{mnras}

\usepackage{newtxtext,newtxmath}

\usepackage[T1]{fontenc}

\DeclareRobustCommand{\VAN}[3]{#2}
\let\VANthebibliography\thebibliography
\def\thebibliography{\DeclareRobustCommand{\VAN}[3]{##3}\VANthebibliography}

\usepackage{graphicx}	
\usepackage{amsmath}	
\usepackage{natbib}
\usepackage{color}
\usepackage{tabularx}
\usepackage{longtable}
\usepackage{bm}
\usepackage{threeparttable}
\usepackage{CJK}

\newcommand{\thetae}{\theta_{\rm E}}

\newcommand{\pie}{\pi_{\rm E}}

\newcommand{\te}{t_{\rm E}}
\newcommand{\eventa}{OGLE-2018-BLG-0383}
\newcommand{\eventb}{KMT-2018-BLG-0998}
\newcommand{\eventc}{OGLE-2018-BLG-0271}

\newcommand{\hjd}{{\rm HJD}^{\prime}}

\title[One Wide-Orbit Planet And Two Stellar Binaries]{Systematic KMTNet Planetary Anomaly Search, Paper III: One Wide-Orbit Planet And Two Stellar Binaries}

\author[Wang et al.]{
Hanyue Wang (王涵悦)$^{1,2}$,
Weicheng Zang (臧伟呈)$^{1,A,C}$, 
Wei Zhu (祝伟)$^{1,C}$\thanks{E-mail: weizhu@mail.tsinghua.edu.cn},
Kyu-Ha Hwang$^{3,A}$, 
\newauthor
Andrzej Udalski$^{4,B}$, 
Andrew Gould$^{5,6,A}$, 
Cheongho Han$^{7,A}$, 
Michael D. Albrow$^{8,A}$, 
Sun-Ju Chung$^{3,9,A}$,
\newauthor
Youn Kil Jung$^{3,A}$,
Doeon Kim$^{7,A}$,
Yoon-Hyun Ryu$^{3,A}$,
In-Gu Shin$^{3,A}$,
Yossi Shvartzvald$^{10,A}$,
Jennifer C. Yee$^{11,A}$,
\newauthor
Sang-Mok Cha$^{3,12,A}$,
Dong-Jin Kim$^{3,A}$,
Hyoun-Woo Kim$^{3,A}$,
Seung-Lee Kim$^{3,9,A}$,
Chung-Uk Lee$^{3,A}$,
\newauthor
Dong-Joo Lee$^{3,A}$,
Yongseok Lee$^{3,12,A}$,
Byeong-Gon Park$^{3,9,A}$,
Richard W. Pogge$^{6,A}$,
Radoslaw Poleski$^{4,B}$,
\newauthor
Przemek Mr\'{o}z$^{4,13,B}$,
Jan Skowron$^{4,B}$,
Micha{\l}~K. Szyma\'{n}ski$^{4,B}$,
Igor Soszy\'{n}ski$^{4,B}$,
Pawe{\l} Pietrukowicz$^{4,B}$,
\newauthor
Szymon Koz{\l}owski$^{4,B}$,
Krzysztof Ulaczyk$^{14,B}$,
Krzysztof A. Rybicki$^{4,B}$,
Patryk Iwanek$^{4,B}$,
Marcin Wrona$^{4,B}$,
\newauthor
Mariusz Gromadzki$^{4,B}$,
Hongjing Yang (杨弘靖)$^{1,C}$,
Shude Mao (毛淑德)$^{1,15,C}$,
Xiangyu Zhang (张翔宇)$^{1,C}$
\\
$^{1}$Department of Astronomy, Tsinghua University, Beijing 100084, China\\
$^{2}$Harvard College, Harvard University, MA 02138, USA\\
$^{3}$Korea Astronomy and Space Science Institute, Daejon 34055, Republic of Korea\\
$^{4}$Astronomical Observatory, University of Warsaw, Al. Ujazdowskie 4, 00-478 Warszawa, Poland\\
$^{5}$Max-Planck-Institute for Astronomy, K\"onigstuhl 17, 69117 Heidelberg, Germany\\
$^{6}$Department of Astronomy, Ohio State University, 140 W. 18th Ave., Columbus, OH 43210, USA\\
$^{7}$Department of Physics, Chungbuk National University, Cheongju 28644, Republic of Korea\\
$^{8}$University of Canterbury, Department of Physics and Astronomy, Private Bag 4800, Christchurch 8020, New Zealand\\
$^{9}$University of Science and Technology, Korea, (UST), 217 Gajeong-ro Yuseong-gu, Daejeon 34113, Republic of Korea\\
$^{10}$Department of Particle Physics and Astrophysics, Weizmann Institute of Science, Rehovot 76100, Israel\\
$^{11}$Center for Astrophysics $|$ Harvard \& Smithsonian, 60 Garden St., Cambridge, MA 02138, USA\\
$^{12}$School of Space Research, Kyung Hee University, Yongin, Kyeonggi 17104, Republic of Korea\\
$^{13}$Division of Physics, Mathematics, and Astronomy, California Institute of Technology, Pasadena, CA 91125, USA\\
$^{14}$Department of Physics, University of Warwick, Gibbet Hill Road, Coventry, CV4~7AL,~UK\\
$^{15}$National Astronomical Observatories, Chinese Academy of Sciences, Beijing 100101, China\\
$^{A}$The KMTNet Collaboration\\
$^{B}$The OGLE Collaboration\\
$^{C}$The Tsinghua Microlensing Team\\
}

\pubyear{2021}

\begin{document}
\begin{CJK*}{UTF8}{gbsn}
\label{firstpage}
\pagerange{\pageref{firstpage}--\pageref{lastpage}}
\maketitle

\begin{abstract}
Only a few wide-orbit planets around old stars have been detected, which limits our statistical understanding of this planet population. Following the systematic search for planetary anomalies in microlensing events found by the Korea Microlensing Telescope Network (KMTNet), we present the discovery and analysis of three events that were initially thought to contain wide-orbit planets. The anomalous feature in the light curve of OGLE-2018-BLG-0383 is caused by a planet with mass ratio $q=2.1\times 10^{-4}$ and a projected separation $s=2.45$. This makes it the lowest mass-ratio microlensing planet at such wide orbits. The other two events, KMT-2018-BLG-0998 and OGLE-2018-BLG-0271, are shown to be stellar binaries ($q>0.1$) with rather close ($s<1$) separations.
We briefly discuss the properties of known wide-orbit microlensing planets and show that the survey observations are crucial in discovering and further statistically constraining such a planet population.
\end{abstract}




\section{Introduction} \label{sec:intro}
Thousands of exoplanets have been detected since the first detection of an exoplanet around a Sun-like star \citep{Mayor:1995}, thanks to the joint effort of many different detection techniques. The majority of the known detections have relatively close-in orbits ($\lesssim1$ AU) and/or large masses ($\gtrsim M_{\rm J}$), and the planets at wide separations---especially those with small masses---remain poorly explored (see recent reviews by \citealt{WinnFabrycky:2015} and \citealt{ZhuDong:2021}).

Perhaps the most efficient method to detect low-mass, wide-orbit planets is gravitational microlensing. Microlensing is most sensitive to planets around the Einstein ring radius
\begin{equation}
    \theta_{\rm E} \equiv \sqrt{\kappa M_{\rm L} \pi_{\rm rel}}; \qquad \kappa \equiv \frac{4G}{c^2 \rm AU} = 8.14 \frac{\rm mas}{M_\odot},
\end{equation}
where $\pi_{\rm rel}$ is the relative parallax between the lens and the source, $M_{\rm L}$ is the mass of the lens \citep{Gould2000}. For typical Galactic events with a lens distance $D_{\rm L}$, the physical Einstein ring radius, $r_{\rm E} \equiv D_{\rm L}\theta_{\rm E}$, corresponds to a few AU \citep{Shude1991,Andy1992}. Planets at such wide separations have long orbital periods and introduce small reflex motions on their hosts, making other methods such as radial velocity very inefficient. So far microlensing has detected over 100 planets, the majority of which have the planet-star projected separation of a factor of two within the Einstein ring radius (see Figure 1 of \cite{OB191053} for an illustration). 

At even larger separations, the lensing signals due to the planet and its host star are largely decoupled, resulting in a reduced sensitivity to planet detections. Although high-magnification events are sensitive to wide-orbit planets via the central caustic \citep{Griest1998}, it is usually difficult to unambiguously determine the host--planet separation due to the close/wide degeneracy \citep{Griest1998,Dominik1999}. Nevertheless, microlensing has yielded a few detections at such wide separations. For example, \citet{OB08092} reported the discovery of a microlensing planet with a projected separation of $5.26 \pm 0.11$ times the Einstein ring radius and the planet-to-star mass ratio $q = (2.41 \pm 0.45) \times 10^{-4}$. For the inferred lens host mass of $0.7\,M_\odot$, these correspond to an orbital separation of $\sim 19$\,AU and a planet mass of $\sim 60\,M_\oplus$, respectively. The planet is thus an ice giant in a Uranus-like orbit \citep{OB08092}. \citet{Poleski:2021} conducted a systematic search for wide-orbit planets in nearly 20 years of microlensing data collected by the Optical Gravitational Lensing Experiment \citep[OGLE,][]{Udalski1994} and found six in the planetary mass regime (mass ratio $q$ in the range of $10^{-4}$--0.033) with projected separation beyond twice the Einstein ring radius. Using the detection efficiency estimated from their extensive simulations, the authors concluded that wide-orbit exoplanets are common, with each microlensing star hosting $\sim 1.4$ such ``ice giants''. The derived rate bears a large statistical uncertainty, primarily due to the limited size of the planet sample.

In this work, we report the detections of wide-orbit planets from the Korea Microlensing Telescope Network \citep[KMTNet,][]{KMT2016}. The three events reported here were discovered in the KMTNet AnomalyFinder algorithm for planet anomalies \citep{OB191053} in the 2018 high-cadence events ($\Gamma_{\rm K} \geq 2~{\rm hr}^{-1}$, \citealt{KB190253}) and first classified as (candidate) planetary events with separations beyond roughly twice the Einstein ring radius, although later detailed modelings revealed that two of them are in fact stellar binaries with close orbits. We describe the observations of the reported events in Section~\ref{sec:observation}, explain our analysis of the microlensing light curves in Section~\ref{sec:model}, and derive physical parameters of the lens systems in Section~\ref{sec:lens}. A discussion of our results is provided in Section~\ref{sec:discussion}.

\section{Observations}\label{sec:observation}

The two lensing events OGLE-2018-BLG-0383/KMT-2018-BLG-0900 and OGLE-2018-BLG-0271/KMT-2018-BLG-0879 were both first detected by the Early Warning System \citep{Udalski1994, Udalski2003} of the fourth phase of OGLE \citep{OGLEIV} and later found by applying the KMTNet EventFinder algorithm \citep{Kim2018a} to all the data collected during the 2018 season. Hereafter, we designate these events by the OGLE names because they made the discoveries first. The third event, \eventb, was detected solely by the KMTNet survey. 

The OGLE data were taken using the 1.3 m Warsaw Telescope equipped with a 1.4 ${\rm deg}^2$ FOV mosaic CCD camera at the Las Campanas Observatory in Chile. \eventa\ and \eventc\ lie in the OGLE BLG500 and BLG504 fields, respectively, with a cadence of $\Gamma_{\rm O} = 1~{\rm hr}^{-1}$. All three events were located in two overlapping KMTNet fields (BLG02 and BLG42), with a combined cadence of $\Gamma_{\rm K} = 4~{\rm hr}^{-1}$. KMTNet consists of three identical 1.6\,m telescopes equipped with 4 ${\rm deg}^2$ FOV cameras at the Cerro Tololo Inter-American Observatory (CTIO) in Chile (KMTC), the South African Astronomical Observatory (SAAO) in South Africa (KMTS), and the Siding Spring Observatory (SSO) in Australia (KMTA).  For both OGLE and KMTNet groups, the great majority of observations were taken in the $I$ band, although $V$ band observations were also taken for the purpose to determine the color of source stars. This work makes use of the $V$ band data from KMTC, which were taken once every ten $I$ band observations.
We summarize in Table~\ref{tab:event_info} the event name, observational cadence, and equatorial and galactic coordinates of the individual events. 

The data used in the light curve analysis were reduced using variants of difference image analysis (DIA, \citealt{Tomaney1996,Alard1998}): \citet{Wozniak2000} for the OGLE data and \citet{pysis} for the KMTNet data. For the KMTC data of each event, we conduct pyDIA photometry\footnote{MichaelDAlbrow/pyDIA: Initial Release on GitHub, doi:10.5281/zenodo.268049} to measure the source color. 

\begin{table}
    \renewcommand\arraystretch{1.5}
    \centering
    \caption{Event Names, Locations and Cadences for the three events}
    \begin{tabular}{c c c c}
    \hline
    \hline
    Name & \eventa & \eventb & \eventc \\
    & /KMT-2018-BLG-0900 & & /KMT-2018-BLG-0879\\
    \hline
    ${\rm RA}_{\rm J2000}$ & 17:54:43.38 & 17:50:59.89 & 17:56:42.25 \\
    ${\rm Decl.}_{\rm J2000}$ & $-$28:44:21.4 & $-$29:32:06.50 & $-$28:23:24.3 \\
    $\ell$ & $1.19$ & $0.09$ & $1.71$ \\
    $b$ & $-1.61$ & $-1.31$ & $-1.81$ \\
    $(\Gamma_{\rm O}, \Gamma_{\rm K})$ $({\rm hr}^{-1})$ & (1, 4) & (0, 4) & (1, 4)\\
    \hline
    \end{tabular}
    \label{tab:event_info}
\end{table}

\section{Light-curve Analysis}\label{sec:model}
\subsection{\rm Preamble} \label{sec:preamble}

All three events show one or two additional bumps to an otherwise normal \cite{Paczynski1986} light curve (single lens and single source, or 1L1S). In such cases the 2L1S (binary lens, single source) parameters can often be inferred based on the morphology of the light curves without extensive numerical searches. The standard 1L1S light curve can be characterized by three parameters: $t_{0}$, the time of the closest lens-source alignment; $u_{0}$, the distance between the lens and the source at the closest alignment in units of the angular Einstein radius, $\thetae$; and $\te$, the timescale it takes to cross the unit Einstein radius
\begin{equation}
    \te \equiv \frac{\thetae}{\mu_{\rm rel}} .
\end{equation}
Here $\mu_{\rm rel}$ is the relative proper motion between the lens and the source. In the case of 2L1S, the center of mass of the binary is used in the definition of $t_0$ and $u_0$. For each data set, we also introduce two flux parameters ($f_{\rm S}$, $f_{\rm B}$) to represent the baseline flux of the source star and any additional blend flux. 

We fit the 1L1S model excluding data around bumps to obtain $(t_{0}, u_{0}, \te)$. The location of a bump can be  estimated at $t_{\rm anom}$ by eye, leading to the offset from the peak $\tau_{\rm anom}$ and the offset from the host $u_{\rm anom}$, both in units of $\thetae$,
\begin{equation}
\tau_{\rm anom}=\frac{t_{\rm anom}-t_{0}}{\te};\qquad u_{\rm anom}=\sqrt{u_{0}^{2}+\tau_{\rm anom}^{2}}. 
\end{equation}
These lead to two 2L1S parameters, $(s, \alpha)$, where $s$ is the projected separation between the binary components normalized to $\thetae$, and $\alpha$ is the angle of source trajectory with respect to the binary axis \citep{Andy1992}, for which the lens mass center is to the right of source forward direction,

\begin{equation}
 \left|\alpha\right|=\left|\sin^{-1} \frac{u_{0}}{u_{\rm anom}}\right|; \qquad s_{\pm}\sim\frac{\sqrt{u_{\rm anom}^{2}+4}\pm u_{\rm anom}}{2}.
\end{equation}
If the source interacts with the minor-image (triangular) planetary caustics, we take $s\simeq s_{-}$, where as if the source interacts with the major-image (quadrilateral) planetary caustic, we expect $s\simeq s_{+}$. The estimates for the remaining two 2L1S parameters, $(q, \rho)$, where $\rho$ is the source radius normalized by $\thetae$, vary in different caustic-passing regimes, and they will be discussed later for individual events separately.

In order to cover all the possible 2L1S models, we also conduct a grid search over the parameter plane $(\log s, \log q, \alpha, \log \rho)$ for each event. The grid consists of 21 values equally spaced between $-1\leq \log s \leq 1$, 51 values equally spaced between $-5 \leq \log q \leq 0$, 10 values equally spaced between $0^{\circ} \leq \alpha < 360^{\circ}$, and 5 values equally spaced between $-3 \leq \log \rho \leq -1$.  For each set of $(\log s, \log q, \alpha, \log \rho)$, we fix $\log q$, $\log s$ and let the other parameters $(t_{0}, u_{0}, \te, \rho, \alpha)$ vary. We use the advanced contour integration code \texttt{VBBinaryLensing} \citep{Bozza2010,Bozza2018} to calculate the magnification of the 2L1S model, and identify the best-fit solution via the Markov chain Monte Carlo (MCMC) method \citep[\texttt{emcee},][]{emcee}.

A short-lived bump on an otherwise normal 1L1S curve can also be caused by the introduction of a second source (single-lens binary-source model, or 1L2S; \citealt{Gaudi1998}), which compared to the primary source is much fainter and passes closer to the lens. 
The total magnification of a 1L2S model is the superposition of two point-lens events, 
\begin{equation}
    A_\lambda = \frac{A_1f_{1,\lambda}+A_2f_{2,\lambda}}{f_{1,\lambda}+f_{2,\lambda}} = \frac{A_{1}+q_{f,\lambda}A_{2}}{1+q_{f,\lambda}}; \quad     q_{f,\lambda} \equiv \frac{f_{2,\lambda}}{f_{1,\lambda}}.
\end{equation}
Here $A_\lambda$ is total magnification, and $f_{{\rm i},\lambda}$ is the baseline flux at wavelength $\lambda$ of each source, with $i=1$ and $2$ corresponding to the primary and the secondary sources, respectively. 
We search for the best-fit 1L2S model for \eventa\ and \eventc. \eventb\ has clear caustic-crossing features that cannot be reproduced by 1L2S models, and thus we do not attempt to perform the 1L2S modeling.

For each event, we also check whether the fit can be further improved after the inclusion of high-order effects. The first is the annual parallax effect \citep{Gould1992,Gould2000,Gouldpies2004}, in which Earth's acceleration around the Sun introduces deviation from rectilinear motion between the lens and the source. The parallax effect is described by two parameters, $\pi_{\rm E,N}$ and $\pi_{\rm E,E}$, which are the north and east component of the microlensing parallax vector $\bm{\pie}$ in equatorial coordinates
\begin{equation}
    \bm{\pie} \equiv \frac{\pi_{\rm rel}}{\thetae} \frac{\bm{\mu_{\rm rel}}}{\mu_{\rm rel}} .
\end{equation}
The second effect is the lens orbital motion \citep{MB09387, OB09020}, which is usually described by two parameters ($ds/dt, d\alpha/dt$), the instantaneous changes in the separation and orientation of the two components defined at $t_0$. We restrict the MCMC trials to $\beta < 0.8$, where $\beta$ is the absolute value of the ratio of projected kinetic to potential energy \citep{An2002,OB050071D},
\begin{equation}
    \beta \equiv \left| \frac{{\rm KE}_{\perp}}{{\rm PE}_{\perp}} \right| = \frac{\kappa M_{\odot} {\rm yr}^2}{8\pi^2}\frac{\pie}{\thetae}\gamma^2\left(\frac{s}{\pie + \pi_{\rm S}/\thetae}\right)^3;\quad \vec{\gamma} \equiv \left(\frac{ds/dt}{s}, \frac{d\alpha}{dt}\right),
\end{equation}
where $\pi_{\rm S}$ is the source parallax.

\subsection{\rm OGLE-2018-BLG-0383}
\begin{figure} 
    \centering
    \includegraphics[width=0.70\columnwidth]{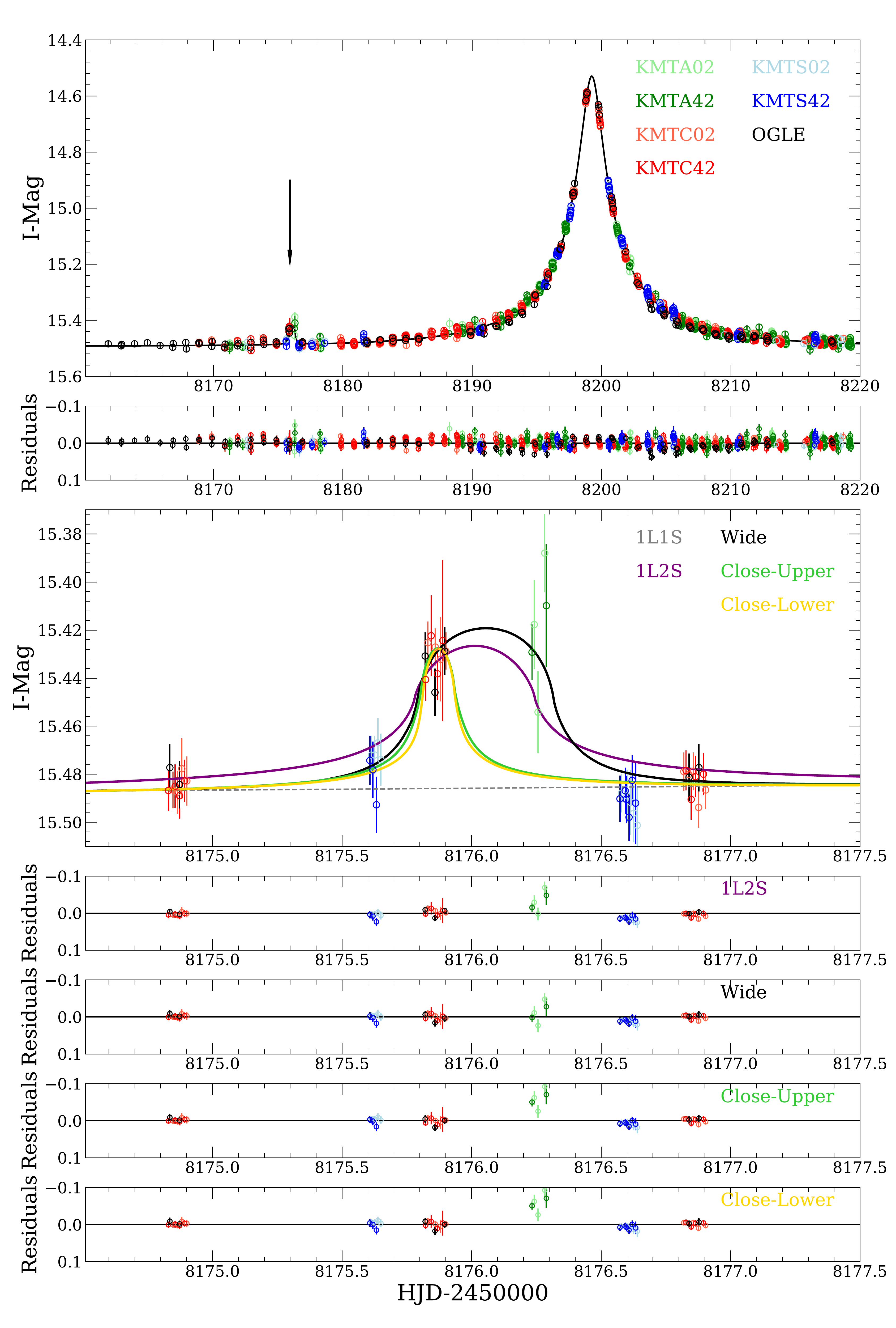}
    \caption{The observed data and models for \eventa. The open circles with different colors are data points for different data sets. The solid lines with different colors represent different models, and the grey dashed line represents the best-fit 1L1S model. In the top panel, the black arrow indicates the position of the planetary signal. The bottom five panels show a close-up of the planetary signal and the residuals to different models.}
    \label{lc_0383}
\end{figure}

Figure \ref{lc_0383} shows the observed light curve of \eventa. There is a $\Delta I \sim 0.07$ mag bump during $8175.5 \lesssim \hjd \lesssim 8176.5$ (${\rm HJD}^{\prime} \equiv {\rm HJD} - 2450000$). The bump appears in multiple data sets (KMTC, KMTA, and OGLE) and all the data points were taken under seeings below or close to the median seeing of the corresponding site. Therefore, the bump is of astrophysical origin. 

\subsubsection{Heuristic Analysis}
We first fit the 1L1S model excluding the data around the small bump and obtain:
\begin{equation}
    (t_0, u_0, \te) = (8199.2, 0.071, 11.3~{\rm days}),
\end{equation}
which leads to
\begin{equation}
 \tau_{\rm anom} = \frac{t_{\rm anom} - t_0}{\te} = -2.04; \qquad u_{\rm anom} = \sqrt{u_0^2 + \tau_{\rm anom}^2} = 2.05; \qquad  |\alpha| = \sin^{-1}\frac{u_0}{u_{\rm anom}} = 1.98^\circ.
\end{equation}
Then, the position of planetary caustic is:
\begin{equation}
    s_{+} \sim \frac{\sqrt{u_{\rm anom}^2 + 4} + u_{\rm anom}}{2} = 2.46; \qquad s_{-} \sim \frac{\sqrt{u_{\rm anom}^2 + 4} - u_{\rm anom}}{2} = 0.41.
\end{equation}
Because the bump exhibits strong finite source effects \citep{1994ApJ...421L..75G,Shude1994,Nemiroff1994}, we expect that a large source envelops a small caustic. \cite{Gould1997} showed that for the case of $s_{+}$, the excess magnification
\begin{equation}
    \Delta A=\frac{2q}{\rho^{2}}.
\end{equation}
Here $\rho$ can be estimated from the duration of the full width half maximum (FWHM) of the bump, $t_{\rm fwhm} \sim 0.55$ days,
\begin{equation}
    \rho \sim \frac{t_{\rm fwhm}}{2\te} \sim 0.024.
\end{equation}
The excess flux of the bump can be read off the light curve, which, combined with $I_{\rm S}$ from the 1L1S model, leads to
\begin{equation}
    \Delta A = \frac{10^{-0.4I_{\rm anom,peak}}-10^{-0.4I_{\rm anom,base}}}{10^{-0.4I_{\rm S}}} = 0.61,
\end{equation}
where $I_{\rm anom,peak}=15.42$ and $I_{\rm anom,base}=15.49$. The planet-to-star mass ratio $q$ can then be estimated as
\begin{equation}
     q = \frac{\Delta A \rho^{2}}{2} \sim 1.8 \times 10^{-4} .
\end{equation}

For the case of $s_{-}$, because it contains two triangular planetary caustics, we expect two solutions. Furthermore, \cite{Gould1997} showed that a large source enveloping both small triangular caustics (together with intervening tough) tends to generate nearly zero excess magnifications, contrary to what is seen in this event. Therefore, we expect that the source is close to or smaller than the caustic in the $s_-$ solutions. 

\subsubsection{Numerical Analysis}
\begin{figure}
    \centering
    \includegraphics[width=\columnwidth]{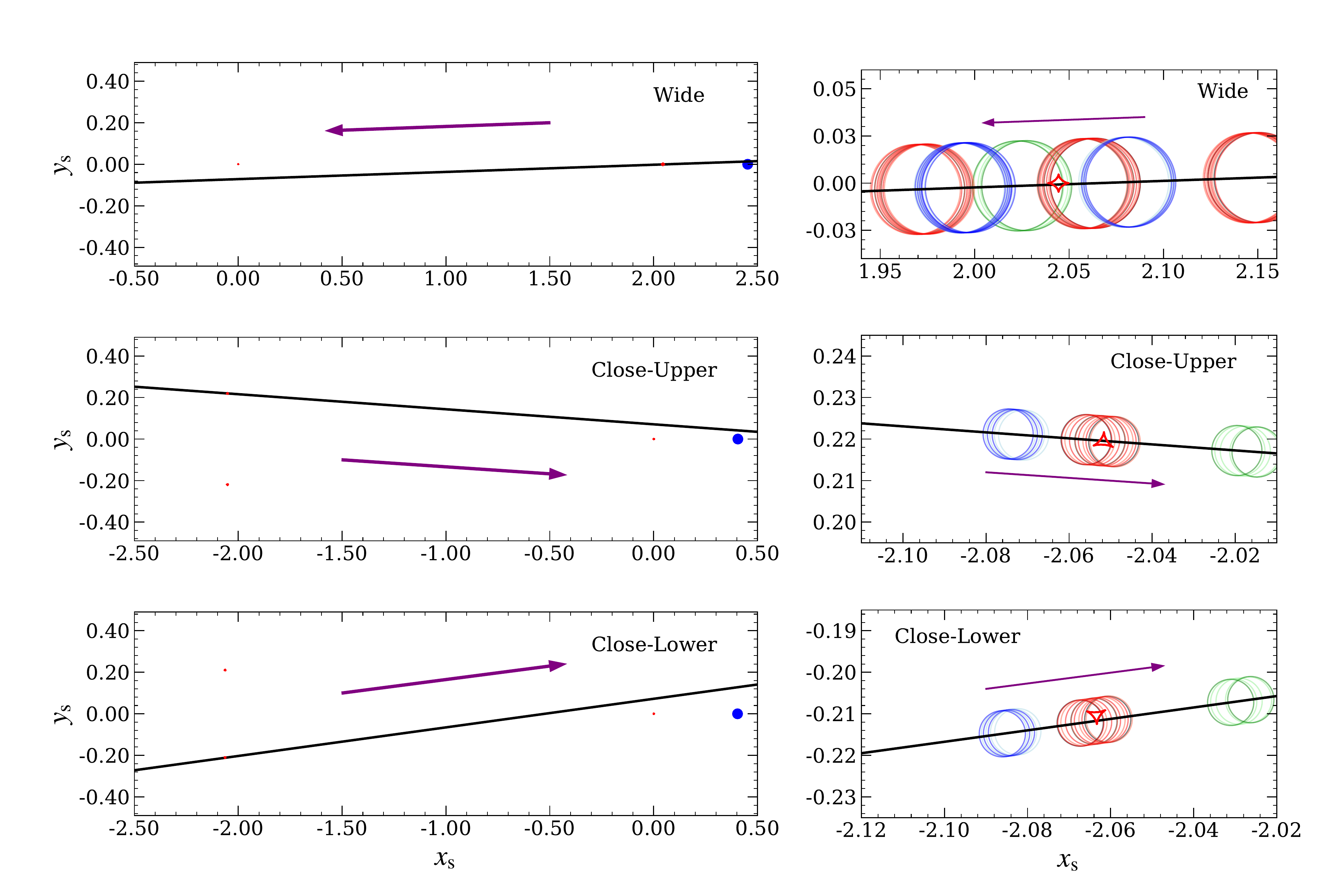}
    \caption{Caustic topologies of the three 2L1S models for \eventa. 
    In each row, the right panel shows the zoomed-in view of the left panel, centering on the caustic-crossing region. We have defined the origin to be the center of mass of the lens system, and the location of the secondary lens is indicated as a filled blue circle. In each panel, the red lines represent the caustic structure, the black solid line represents the source trajectory, the magenta arrow indicates the direction of the source motion, and the open circles with different colors (not shown in the left panels) represent the source location at the times of observation from different telescopes. The radii of the circles represent the normalized source radius $\rho$ of each model. The color scheme is the same as in Figure~\ref{lc_0383}. } 
    \label{cau_0383}
\end{figure}

We conduct a grid search to identify all degenerate solutions, following the description of Section~\ref{sec:preamble}. As expected based on the above analysis, three local minima are identified. For each solution, we then perform MCMC analysis to obtain the best-fit 2L1S parameters. Figure \ref{cau_0383} shows the caustics and source trajectories of the three solutions. As expected, one of the solutions contains a large source crossing a small major-image (quadrilateral) planetary caustic, and the other two have a relatively small source crossing the minor-image (triangular) planetary caustic. We label the three solutions as ``Wide'', ``Close-Upper'', and ``Close-Lower'', respectively. Their best-fit parameters and the associated $68\%$ confidence intervals from the MCMC analyses are given in Table \ref{parm_0383}, and the corresponding light curves are shown in Figure \ref{lc_0383}.  We note that the values of $(s, \alpha, \rho, q)$ from the heuristic analysis are in good agreement with the values from the detailed numerical analyses.

\begin{table*}
    \renewcommand\arraystretch{1.3}
    \centering
    \caption{1L1S, 2L1S and 1L2S model parameters for \eventa. The best-fit solution is highlighted in bold.}
    \begin{tabular}{c|c|c c c|c}
    \hline
    \hline
     & 1L1S & \multicolumn{3}{c|}{2L1S} & 1L2S \\
    \hline 
      & & \textbf{Wide} & Close-Upper & Close-Lower & \\
    \hline
    $\chi^2/dof$ & 2384.4/1874 & \textbf{1870.3/1870} & 1928.8/1870 & 1929.9/1870 & 1905.9/1869 \\
    \hline
    $t_{0,1}$ (${\rm HJD}^{\prime}$) & $8199.244 \pm 0.002$ & $\mathbf{8199.239 \pm 0.003}$ & $8199.247 \pm 0.003$ & $8199.247 \pm 0.002$ & $8199.244 \pm 0.003$ \\
    
    $t_{0,2}$ (${\rm HJD}^{\prime}$) & ... & ... & ... & ... & $8176.022 \pm 0.048$ \\
    
    $u_{0,1}$ & $0.072 \pm 0.001$ & $\mathbf{0.071 \pm 0.001}$ & $0.071 \pm 0.001$ & $ 0.072 \pm 0.001$ & $0.074 \pm 0.005$  \\
    
    $u_{0,2}$ & ... & ... & ... & ... & $0.0007 \pm 0.0018$ \\
    
    $\te$ (days) & $11.15 \pm 0.17$ & $\mathbf{11.35 \pm 0.17}$ & $11.34 \pm 0.17$ & $11.46 \pm 0.21$ & $11.34 \pm 0.22$ \\
    
    $\rho_1$ & ... & $\mathbf{0.0238 \pm 0.0020}$ & $0.0060 \pm 0.0008$ & $0.0056 \pm 0.0007$ & $0.058 \pm 0.021$ \\
    
    $\rho_2$ & ... & ... & ... & ... & $0.0202 \pm 0.0050$ \\
    
    $q_{f,I}$ & ... & ... & ... & ... & $0.0057 \pm 0.0014$ \\
    
    $\alpha$ (deg) & ... & $\mathbf{181.98 \pm 0.17}$ & $355.86 \pm 0.75$ & $7.84 \pm 0.70$ & ... \\
    
    $s$ & ... & $\mathbf{2.453 \pm 0.026}$ & $0.405 \pm 0.004$ & $0.404 \pm 0.005$ & ... \\
    
    $q$ ($10^{-4}$) & ... & $\mathbf{2.14 \pm 0.34}$  & $23.6 \pm 5.9$ & $21.5 \pm 5.2$ & ... \\
    
    $f_{\rm S, OGLE}$ & $1.132 \pm 0.022$ & $\mathbf{1.130 \pm 0.021}$ & $1.127 \pm 0.021$ & $1.117 \pm 0.023$ & $1.119 \pm 0.029$ \\
    
    $f_{\rm B, OGLE}$ & $8.879 \pm 0.020$ & $\mathbf{8.870 \pm 0.019}$ & $8.875 \pm 0.019$ & $8.881 \pm 0.021$ & $8.878 \pm 0.026$ \\
    \hline
    \end{tabular}
    \begin{tablenotes}
    \item All flux values are normalized to a 18th magnitude source, i.e., $I_{\rm S} = 18 - 2.5 \log(f_{\rm S})$.
    \end{tablenotes}
    \label{parm_0383}
\end{table*}

Among all three solutions, the ``wide'' solution provides the best fit to the observed data, especially those around the bump. The ``Close-Upper'' and ``Close-Lower'' solutions are both disfavored by $\Delta\chi^2 > 58$ and cannot fit the five KMTA points at $\hjd \sim 8176.2$. We thus reject the ``Close-Upper'' and ``Close-Lower'' solutions.

We also check the 1L2S model and present its best-fit parameters in Table \ref{parm_0383}. Compared to the 2L1S ``Wide'' model, the 1L2S model has a worse fit by $\Delta\chi^2 = 30.5$, which is already a strong evidence against the 1L2S model.
The 1L2S model is also disfavored for its somewhat nonphysical model parameters. The secondary source has a normalized source radius, $\rho_2 = 0.020 \pm 0.003$. Being $\sim$180 times brighter, the normalized source radius of the primary source should be about one order of magnitude larger and thus $\rho_1 \sim 0.2$. This is inconsistent with $\rho_{1}=0.058 \pm 0.021$ from the light curve analysis.
Furthermore, following the CMD analysis in Section \ref{CMD} and based on the star color of \cite{HSTCMD}, one would get $\thetae \sim 0.02$ mas and $\mu_{\rm rel} \sim 0.6~{\rm mas\,yr^{-1}}$ for the 1L2S model. Lenses with such kinematics are fairly rare according to the standard Galactic model (See Figure 2 of \citealt{Zhu2017spitzer}). Hence, the 1L2S model is also rejected.

The inclusion of the annual parallax and the lens orbital motion effects only improves the fit by $\Delta\chi^2 < 2$. Such an improvement is too small compared to the impact of typical systematics in the data. Furthermore, the inclusion of the parallax effect yields a $1\sigma$ upper limit on $\pie$ of $\sim1.5$, which is too large to be considered physically meaningful. This is expected, given that the event has a short timescale ($\te=11.4\,$days). As the inclusion of the higher-order effects gives statistically similar values for the standard microlensing parameters, we adopt the static binary solution as the final solution.

\medskip

\subsection{\rm KMT-2018-BLG-0998}

\begin{figure}
    \centering
    \includegraphics[width=0.70\columnwidth]{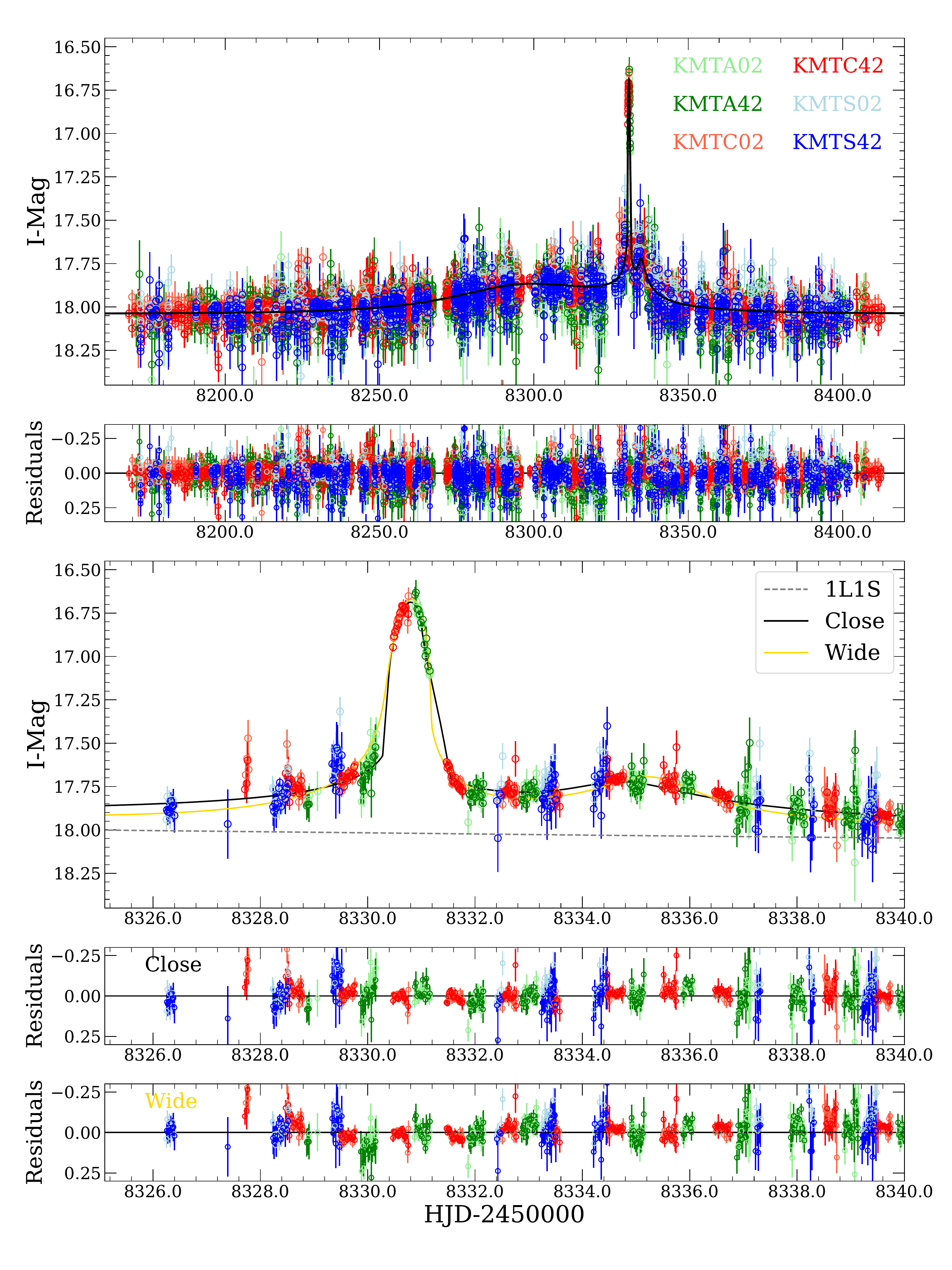}
    \caption{The observed data and models for \eventb. The symbols are similar to those in Figure \ref{lc_0383}. The two bumps in the third panel are produced by the source approaching the two spikes of the caustic. See Figure \ref{cau_0998} for the lensing geometry. }
    \label{lc_0998}
\end{figure}

As shown in Figure~\ref{lc_0998}, the light curve of event KMT-2018-BLG-0998 shows two bumps in addition to the 1L1S model, with both brighter than the primary peak of the 1L1S model. Such features can be produced by the source crossing or approaching the two spikes of the planetary caustic. 

\subsubsection{Heuristic Analysis}
We first fit the 1L1S model excluding data around the two bumps and obtain:
\begin{equation}
    (t_0, u_0, \te) = (8301.3, 1.09, 29.1~{\rm days}) .
\end{equation}
Together with the central time of the planetary anomaly, $t_{\rm anom}\approx8333$, these lead to
\begin{equation}
 \tau_{\rm anom} = \frac{t_{\rm anom} - t_0}{\te} \approx 1.09; \qquad u_{\rm anom} = \sqrt{u_0^2 + \tau_{\rm anom}^2} \approx 1.54; \qquad |\alpha| = \sin^{-1}\frac{u_0}{u_{\rm anom}} \approx 45^\circ
\end{equation}
We then obtain 
\begin{equation}
    s = s_{+} \sim \frac{\sqrt{u_{\rm anom}^2 + 4} + u_{\rm anom}}{2} = 2.03; \qquad s_{-} \sim \frac{\sqrt{u_{\rm anom}^2 + 4} - u_{\rm anom}}{2} = 0.49.
\end{equation}
We can also estimate the size of the source from the first bump, which exhibits strong finite-source effect. The width of this bump is $t_{\rm fwhm} \sim 0.6$ days, and thus
\begin{equation}
    \rho \sim \frac{t_{\rm fwhm}}{2\te} \sim 0.01.
\end{equation}


\begin{table*}
    \renewcommand\arraystretch{1.3}
    \centering
    \caption{2L1S model parameters of \eventb.}
    \begin{tabular}{c|c|c}
    \hline
    \hline
    & Wide & \textbf{Close}\\
    \hline
    $\chi^2/dof$  & 11473.1/11017 & \textbf{11017.1/11017} \\
    \hline
    $t_{0}$ (${\rm HJD}^{\prime}$) & $8304.183 \pm 0.155$ & $\mathbf{8298.226 \pm 0.211}$\\

    $u_{0}$ & $0.930 \pm 0.005$ & $\mathbf{0.979 \pm 0.010}$\\
    
    $\te$ (days) & $31.23 \pm 0.20$ & $\mathbf{30.33 \pm 0.19}$ \\
    
    $\rho$ & $0.0103 \pm 0.0001$ & $\mathbf{0.0094 \pm 0.0001}$\\
    
    $\alpha$ (deg) & $312.45 \pm 0.23$ & $\mathbf{59.65 \pm 1.16}$ \\
    
    $s$ & $1.917\pm 0.003$ & $\mathbf{0.553\pm 0.002}$\\
    
    $q$ & $0.0196 \pm 0.0003$ & $\mathbf{0.601 \pm 0.029}$\\
    
    $f_{\rm S, KMTC02}$ & $0.472 \pm 0.004$ & $\mathbf{0.551 \pm 0.007}$ \\
    
    $f_{\rm B, KMTC02}$ & $0.498 \pm 0.004$ & $\mathbf{0.414 \pm 0.007}$ \\
    \hline
    \end{tabular}
    \label{parm_0998}
\end{table*}

\subsubsection{Numerical Analysis}
\begin{figure}
    \centering
    \includegraphics[width=\columnwidth]{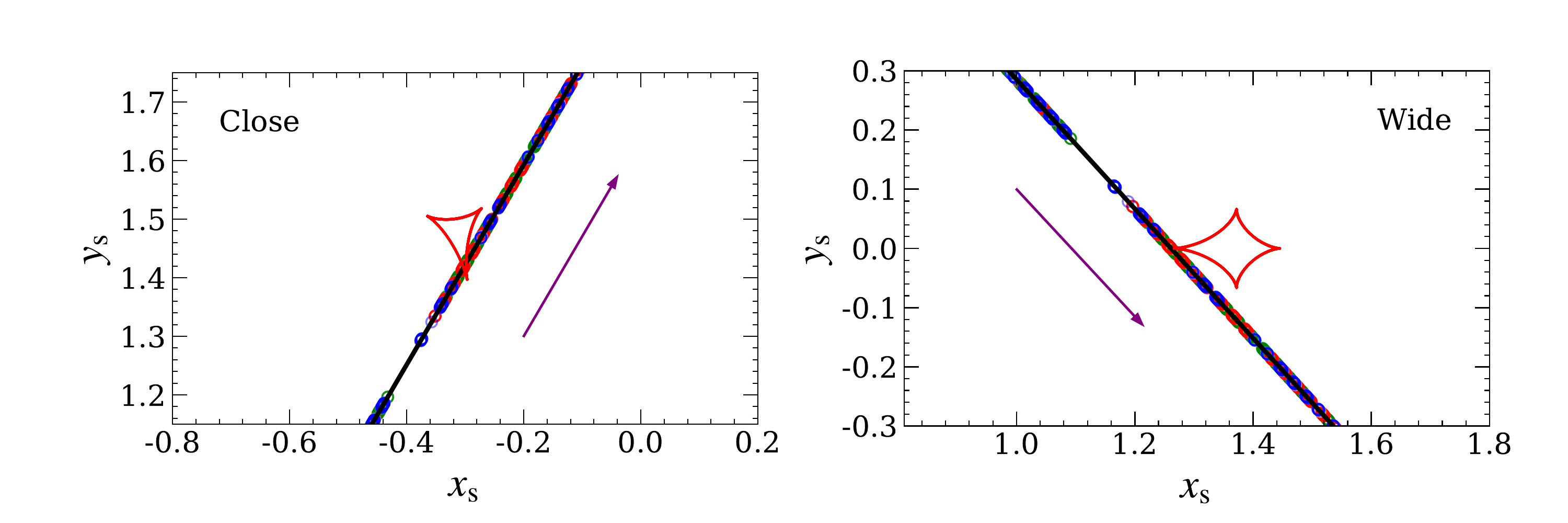}
    \caption{Lensing geometry of \eventb. The symbols are similar to those in Figure \ref{cau_0383}.}
    \label{cau_0998}
\end{figure}

We conduct a grid search that covers both planetary and stellar binary mass ratios and find two local minima in $\chi^2$ in the $q$ vs. $s$ plane. We then perform detailed MCMC modeling to further refine the model parameters. The results are presented in Table~\ref{parm_0998}, and the corresponding caustic structure and source trajectory are shown in Figure~\ref{cau_0998}. We label the $s < 1$ and $s > 1$ solutions as ``Close'' and ``Wide'', respectively. As expected, the two bumps are produced by the source crossing one spike and approaching another spike of the caustic. We find that the ``Close'' solution is favored by $\Delta\chi^2 = 456$ and most of the $\Delta\chi^2$ difference comes from the anomaly region, so we adopt the ``Close'' solution as the final model of this event.  
With $q = 0.6$, this "Close" solution suggests that the lens system is composed of two stars. 

We find that the inclusion of higher-order effects does not change the general interpretation of the lens system. Furthermore, different data sets of this event yield different constraints on the parameters associated with the higher-order effects, suggesting the existence of systematics in some (or all) of the data sets or photometric variability of the target. For the purpose of this work, we will not proceed with further investigations into its origin and simply adopt the parameters of the static 2L1S model as the final solution.

\subsection{\rm OGLE-2018-BLG-0271}

\begin{figure}
    \centering
    \includegraphics[width=0.70\columnwidth]{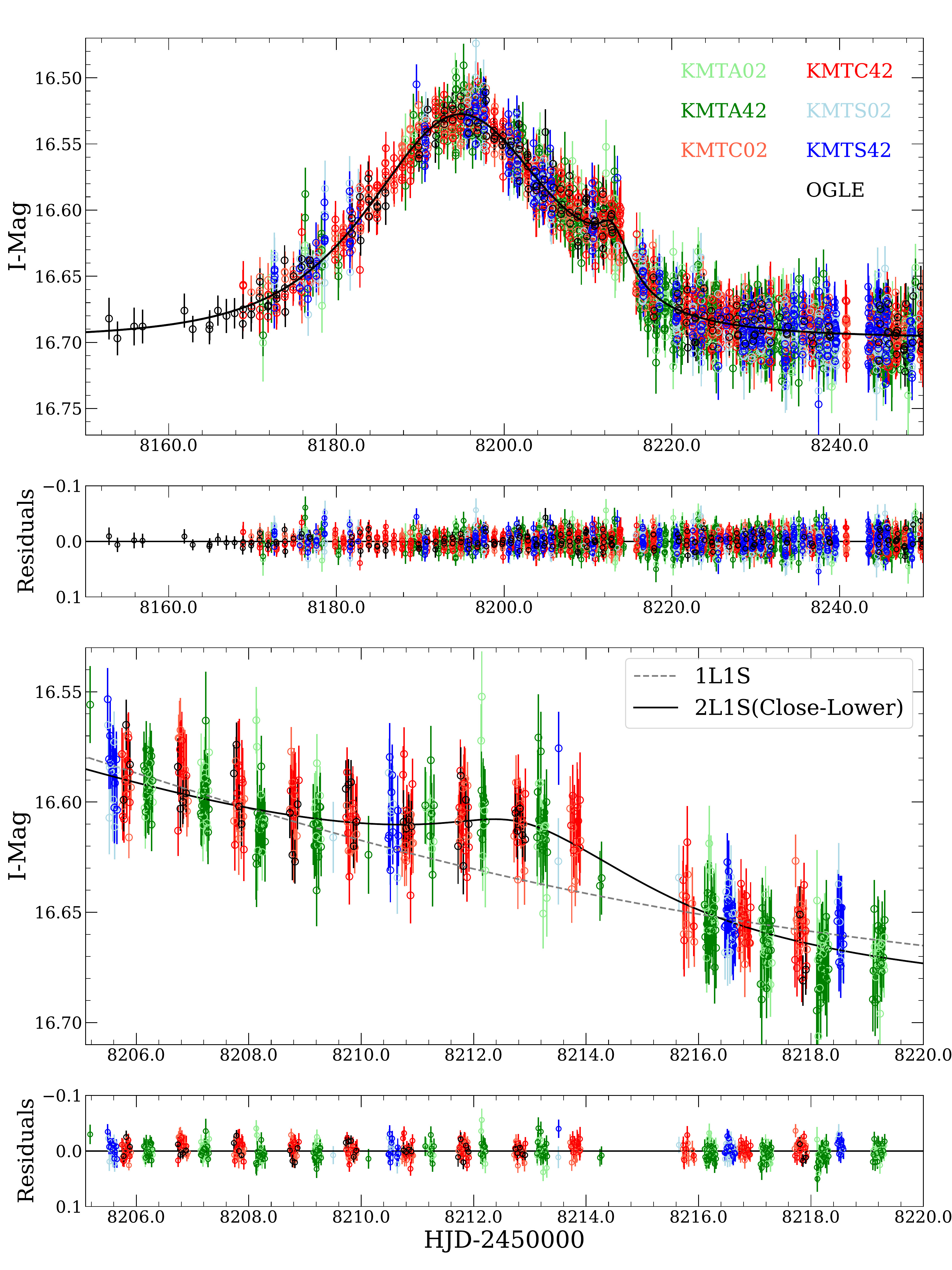}
    \caption{The observed data and the best-fit 1L1S and 2L1S models for \eventc. The symbols are similar to those in Figure \ref{lc_0383}.}
    \label{lc_0271}
\end{figure}

As shown in Figure~\ref{lc_0271}, the light curve of OGLE-2018-BLG-0271 shows a $\sim$6-day bump around $\hjd \sim 8212$. This anomaly is securely detected in all data sets, including OGLE and KMTNet.

\subsubsection{Heuristic Analysis}
The 1L1S model without the data around the bump yields
\begin{equation}
    (t_0, u_0, \te) = (8195.01, 1.42, 10.4~{\rm days}) .
\end{equation}
These lead to
\begin{equation}
 \tau_{\rm anom} = \frac{t_{\rm anom} - t_0}{\te} = 1.63; \qquad u_{\rm anom} = \sqrt{u_0^2 + \tau_{\rm anom}^2} = 2.16; \qquad  |\alpha| = \sin^{-1}\frac{u_0}{u_{\rm anom}} = 41.1^\circ,
\end{equation}
and thus
\begin{equation}
    s_{+} \sim \frac{\sqrt{u_{\rm anom}^2 + 4} + u_{\rm anom}}{2} = 2.55; \qquad s_{-} \sim \frac{\sqrt{u_{\rm anom}^2 + 4} - u_{\rm anom}}{2} = 0.39.
\end{equation}
For $s_{-}$, we again expect two solutions that correspond to two triangular planetary caustics, respectively. For $s_{+}$, because the bump does not exhibit clear finite-source effects, we expect the so-called ``inner/outer degeneracy'', for which the source passes from the inner and outer sides (with respect to the host of the planet) of the major-image planetary caustic, respectively \citep{GaudiGould1997}. 

\begin{table}
    \renewcommand\arraystretch{1.2}
    \centering
    \caption{1L1S, 2L1S and 1L2S model parameters of \eventc. The adopted model is highlighted in bold.}
    \begin{tabular}{{c|c|c c c c|c}}
    \hline
    \hline
     & 1L1S & \multicolumn{4}{c|}{2L1S} & 1L2S \\
    \hline 
      & & Wide-Inner & Wide-Outer & Close-Upper & \textbf{Close-Lower} & \\
    \hline
    $\chi^2/dof$ & 12763.0/12034 & 12084.6/12030 & 12358.1/12030 & 12652.9/12030 & \textbf{12030.4/12030} & 12158.3/12029  \\
    \hline
    $t_{0,1}$ (${\rm HJD}^{\prime}$) & $8195.44 \pm 0.06$ & $8195.56  \pm 0.07$ & $8195.74 \pm 0.07$ & $8196.56 \pm 0.07$ & $\mathbf{8194.21 \pm 0.13}$ & $8194.90 \pm 0.08$ \\
    
    $t_{0,2}$ (${\rm HJD}^{\prime}$) & ... & ... & ...& ... & ... & $8212.22 \pm 0.13$ \\
    
    $u_{0,1}$ & $1.427 \pm 0.026$ & $1.298 \pm 0.035$ & $1.306 \pm 0.025$ & $1.309 \pm 0.004$ & $\mathbf{1.349 \pm 0.075}$ & $1.463 \pm 0.026$ \\
    
    $u_{0,2}$ & ... & ... & ... & ... & ...& $0.157 \pm 0.023$ \\
    
    $t_{\rm E}$ (days) & $11.09 \pm 0.26$ & $10.77 \pm 0.20$ & $10.52 \pm 0.14$ & $8.77 \pm 0.05$ &  $\mathbf{10.74 \pm 0.42}$ & $10.07 \pm 0.16$ \\
    
    $\rho_1$ & ... & $<0.27$ & $<0.31$ & $<0.19$ & $\mathbf{<0.28}$ & $0.51 \pm 0.37$ \\
    
    $\rho_2$ & ... & ... & ... & ... & ... & $0.27 \pm 0.11$  \\
    
    $q_{f,I}$ & ... & ... & ... & ... & ... & $0.0058 \pm 0.0010$ \\
    
    $\alpha$ (deg) & ... & $ 318.67 \pm 0.41$ & $325.78 \pm 1.22$ & $98.78 \pm 0.26$ & $\mathbf{183.92 \pm 4.87}$ & ... \\
    
    $s$ & ... & $3.074 \pm 0.057$ & $1.748 \pm 0.040$ & $0.388 \pm 0.003$ & $\mathbf{0.411 \pm 0.014}$ & ... \\
    
    $q$ & ... & $0.026 \pm 0.003$ & $0.040 \pm 0.007$ & $0.200 \pm 0.004$ & $\mathbf{0.101 \pm 0.024}$ & ... \\
    
    $f_{\rm S, OGLE}$ & $3.72 \pm 0.18$ & $3.61 \pm 0.24$ & $3.48 \pm 0.14$ & $3.60 \pm 0.03$ & $\mathbf{ 3.60 \pm 0.39}$ & $3.99 \pm 0.18$ \\
    
    $f_{\rm B, OGLE}$ & $-0.40 \pm 0.18$ & $-0.29 \pm 0.24$ & $-0.16 \pm 0.14$ & $-0.28 \pm 0.03$ & $\mathbf{-0.28 \pm 0.39}$ & $-0.67 \pm 0.18$ \\
    \hline
    \end{tabular}
    \begin{tablenotes}
    \item The values of $\rho_1$ are their $3\sigma$ ($\Delta\chi^2<9$) upper limits.
    \end{tablenotes}
    \label{parm_0271}
\end{table}

\begin{figure}
    \centering
    \includegraphics[width=0.70\columnwidth]{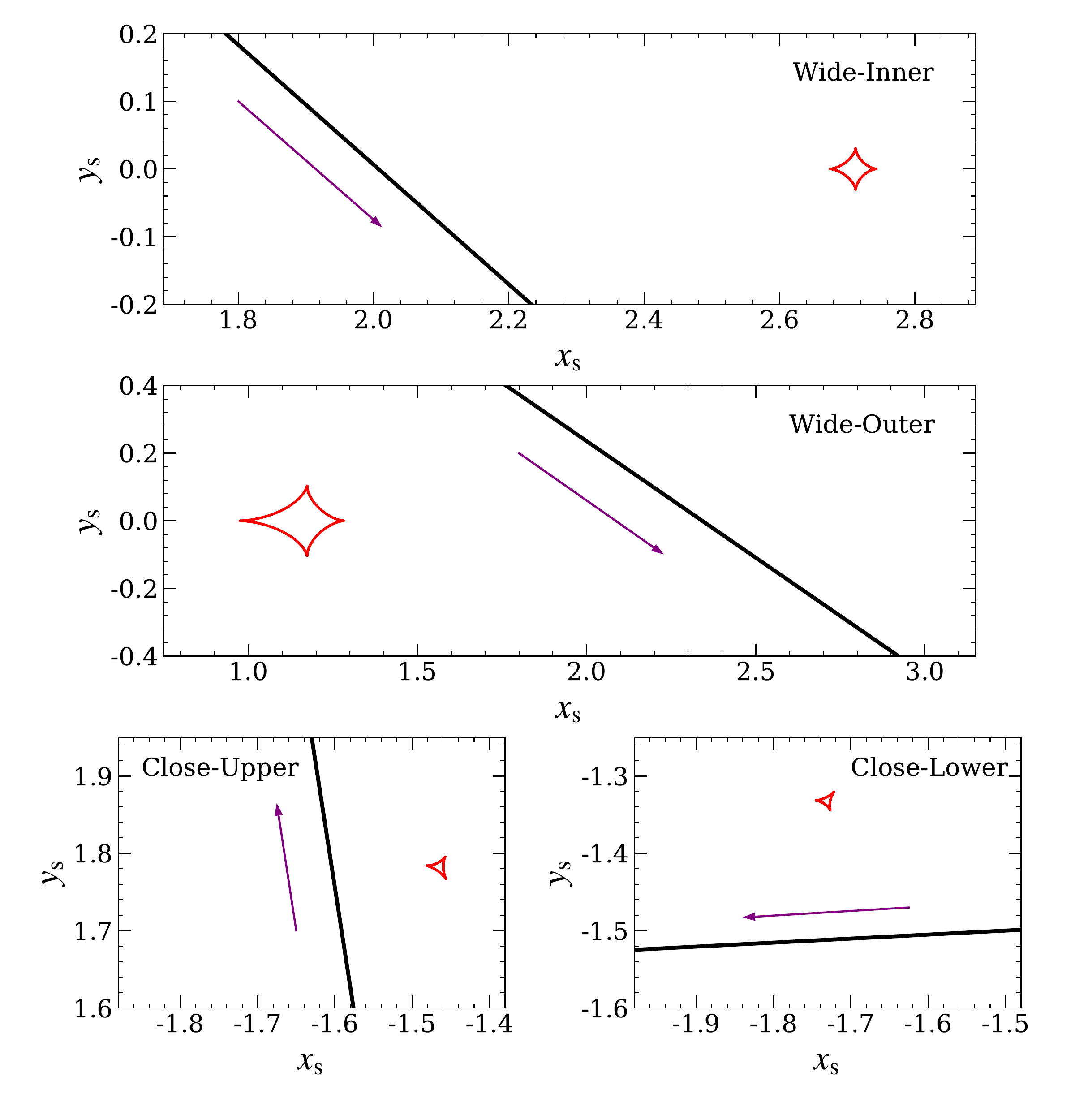}
    \caption{Caustic topologies of the four 2L1S models for \eventc. The symbols are similar to those in Figure \ref{cau_0383}. Because the four models only have upper limits on $\rho$, the source radii are not shown.}
    \label{cau_0271}
\end{figure}

\subsubsection{Numerical Analysis}

\begin{figure}
    \centering
    \includegraphics[width=0.70\columnwidth]{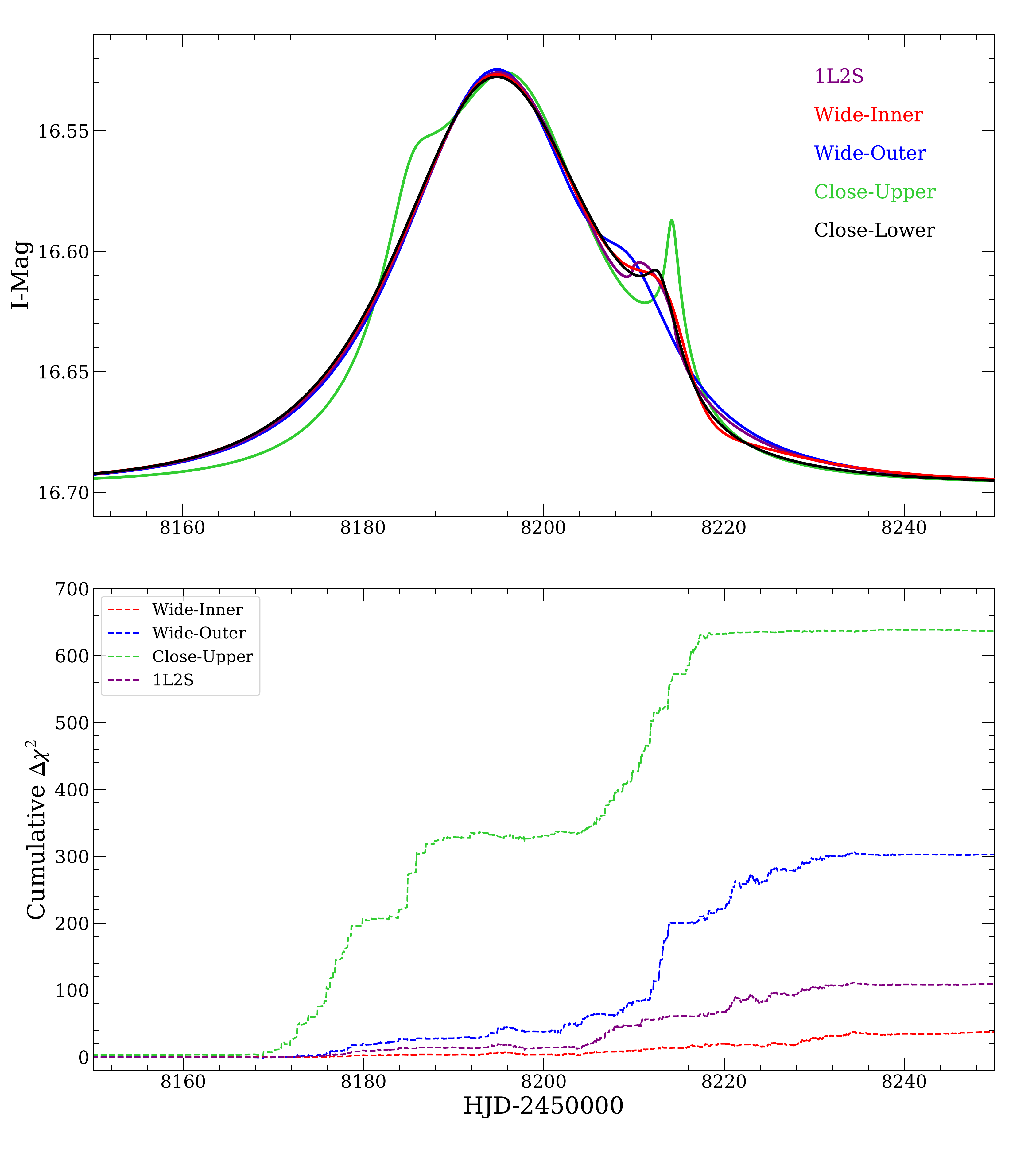}
    \caption{The upper panel shows the best-fit models of the four 2L1S models and the 1L2S model for \eventc. The lower panel shows the cumulative distribution of $\Delta\chi^2$ differences for the three 2L1S models and the 1L2S model relative to the 2L1S ``Close-Lower'' model, which provides the best fit to the observed data.}
    \label{chi2_0271}
\end{figure}

Four local minima are identified in the grid search, which is consistent with the heuristic analysis. Based on the caustic structures and source trajectories (Figure~\ref{cau_0271}), these solutions are labeled as ``Wide-Inner'', ``Wide-Outer'', ``Close-Upper'', and ``Close-Lower'', and their best-fit parameters from the MCMC modelings are presented in Table~\ref{parm_0271} together with the best-fit 1L2S model. We find that the ``Close-Lower'' solution provides the best fit to the observed data, whereas the ``Wide-Inner'', ``Wide-Outer'', ``Close-Upper'' and 1L2S solutions are disfavored by $\Delta\chi^2 > 54$, $328$, $623$, and $128$, respectively. In Figure \ref{chi2_0271}, we show the cumulative $\Delta\chi^2$ distributions of the four solutions relative to the ``Close-Lower'' solution are shown in Figure~\ref{chi2_0271}. The fact that most of the $\Delta\chi^2$ differences come from the anomaly region is a strong indication that the $\Delta\chi^2$ difference is statistically meaningful. We thus adopt the ``Close-Lower'' solution as the final model of this event. This solution has a binary mass ratio with $q \sim 0.1$, suggesting that the companion is probably a brown dwarf or a low-mass star.

High-order effects have also been explored for this event, but it only provides $\Delta\chi^2 \sim 1$ and the $1\sigma$ uncertainty of parallax is $\sim 1$. For reasons similar to the first event, we adopt the 2L1S model without high-order effects.

\section{Physical Parameters}\label{sec:lens}
In principle the mass and distance of the lens system can be determined if both the angular Einstein radius and the microlensing parallax are measured \citep{Gould1992,Gould2000}. 
Unfortunately, the parallax effect is not detected in any of the three events analyzed here, and \eventc\ only has an upper limit on $\rho$ (and thus a lower limit on $\thetae$). Therefore, we rely on the Bayesian analysis to estimate the physical parameters of the lens system.

\subsection{\rm Color Magnitude Diagram}\label{CMD}

\begin{figure}
    \centering
    \includegraphics[width=0.31\textwidth]{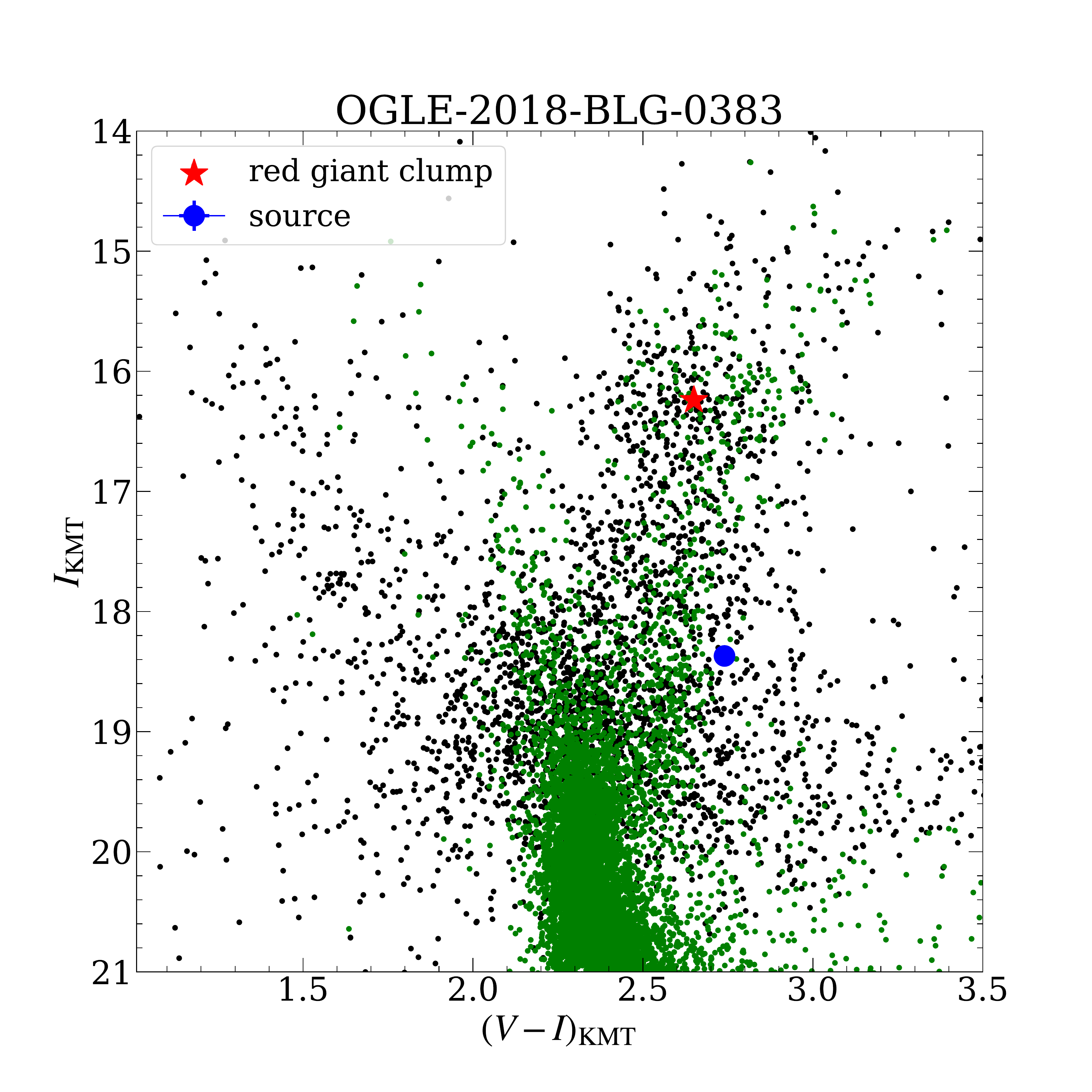}
    \includegraphics[width=0.31\textwidth]{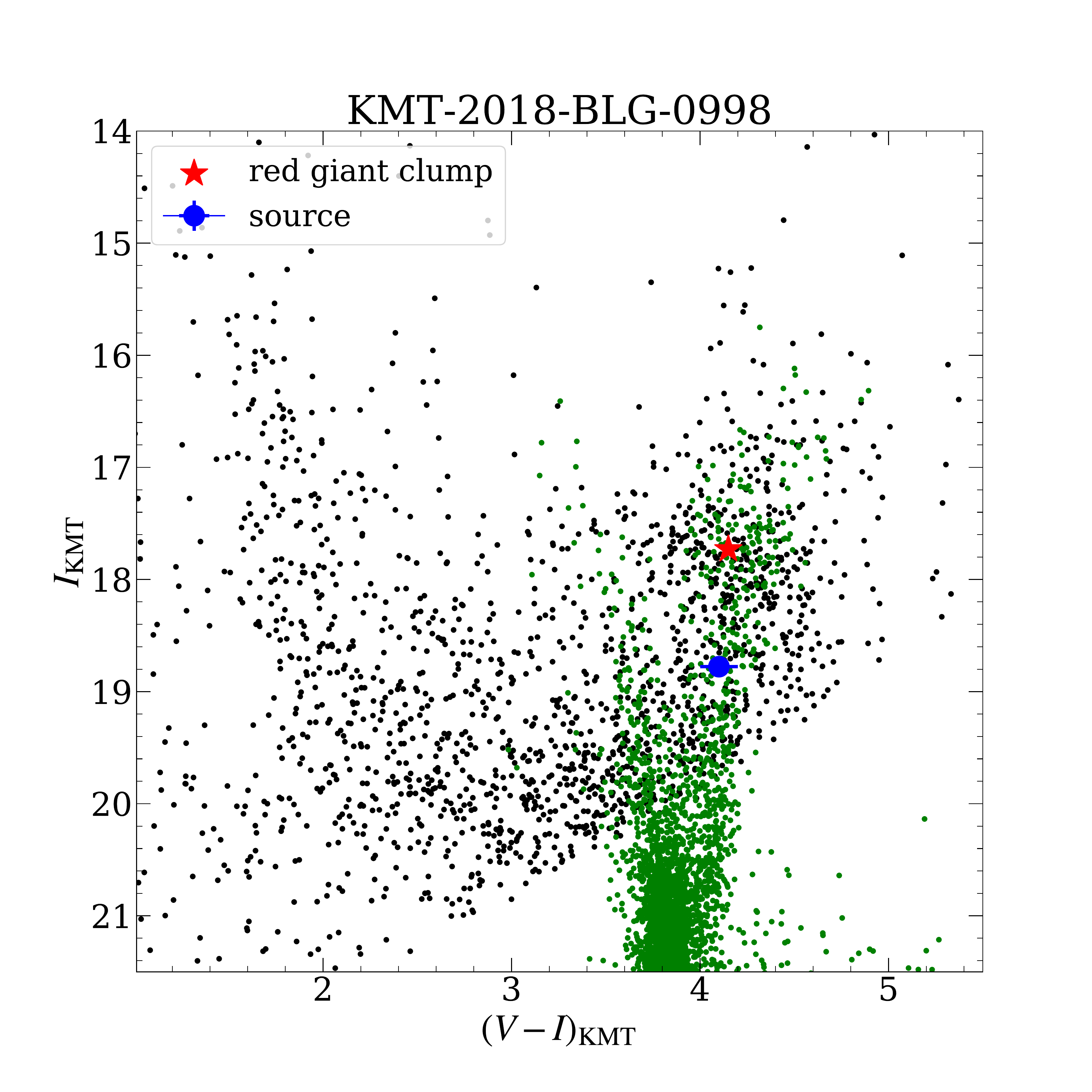}
    \includegraphics[width=0.31\textwidth]{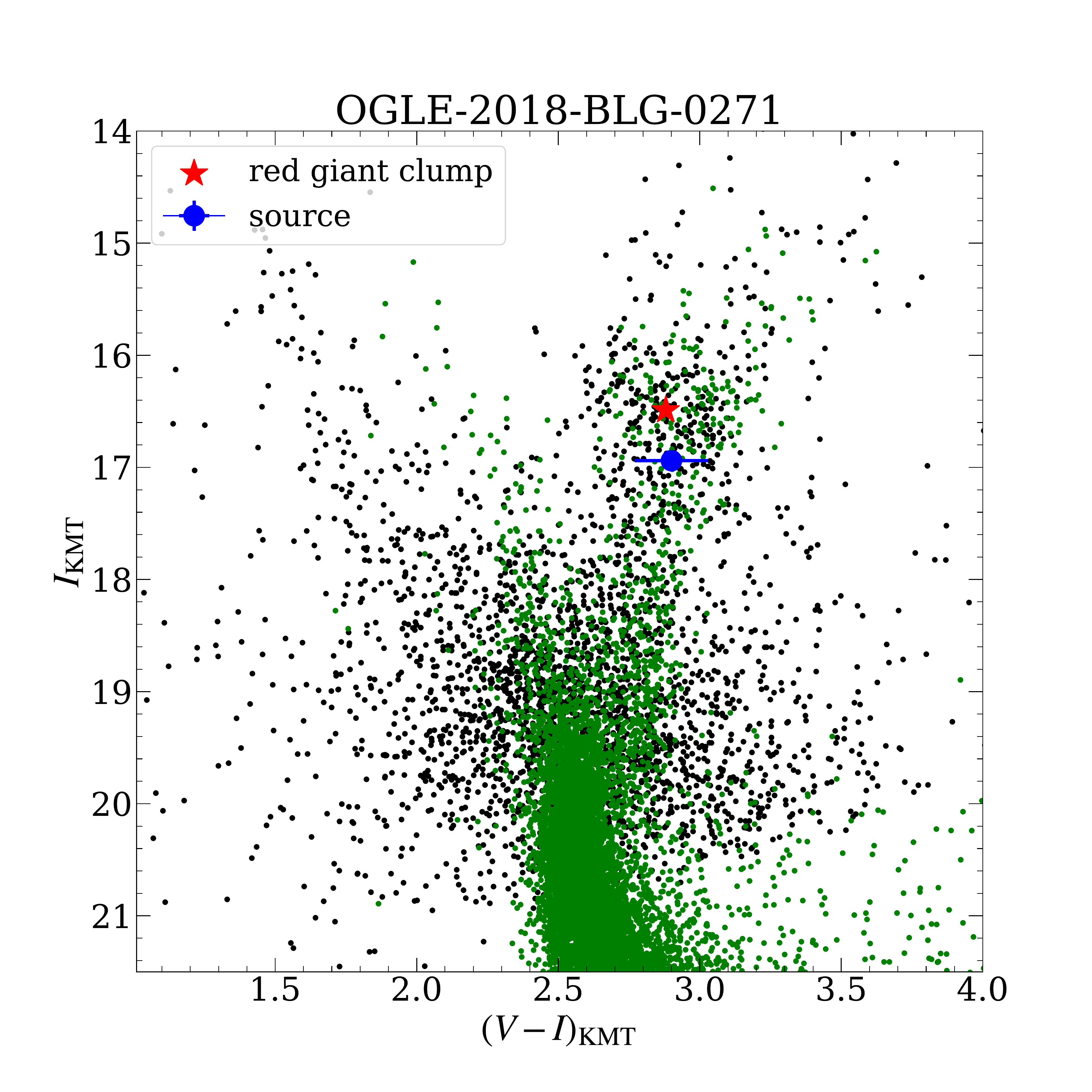}
     \caption{Color-magnitude diagrams (CMDs) of the three microlensing events. Observations from KMTC (black dots) are used to construct these diagrams. For each panel, the red asterisk and blue dot represent the positions of the centroid of the red clump and the source star, respectively. The green dots show the HST CMD of \citet{HSTCMD} whose red-clump centroid has been adjusted to that of KMTC.}
    \label{cmd}
\end{figure}

We first determine the angular radius of the source star, $\theta_\star$, based on a color magnitude diagram (CMD) analysis \citep{Yoo2004}. For each event, we construct a $V-I$ versus $I$ CMD based on the KMTC pyDIA photometry and stars within a $120^{\prime\prime}$ square centered on the event position (see Figure \ref{cmd}). We first estimate the centroid of the red clump as $(V - I, I)_{\rm cl}$ and compare it with the intrinsic centroid of the red clump $(V - I, I)_{\rm cl,0}$.
Here we adopt $(V - I)_{\rm cl,0} = 1.06 \pm 0.03$, with the value and uncertainty taken from \citet{Bensby2013} and \citet{Nataf2016}, respectively. The dereddened magnitudes, $I_{\rm cl,0}$, are taken with an uncertainty of $0.04$ mag from Table~1 of \citet{Nataf2013} at the locations of individual events. These yield the offset
\begin{equation}
    \Delta(V - I, I) = (V - I, I)_{\rm cl} - (V - I, I)_{\rm cl,0}.
\end{equation}
For \eventa, we determine the source color and magnitude $(V - I, I)_{\rm S}$ from a regression of the KMTC pyDIA $V$ versus $I$ flux and the light-curve analysis in Section \ref{sec:model}, respectively. We have also derived the source $V - I$ color from the light-curve analysis and found a consistent result with $1\sigma$. For \eventb\ and \eventc, the source color cannot be determined due to the low S/N of the $V$-band observations, so we follow the method of \cite{MB07192} to estimate the source color from the \emph{Hubble} Space Telescope (HST) CMD of \cite{HSTCMD}. We first calibrate the {\it HST} CMD to the KMTC CMD using their positions of red clump centroid. Then, we estimate the source color by taking the average color of the calibrated {\it HST} stars whose brightness are within $5\sigma$ of the microlensing source star. For each event, we find the dereddened color and magnitude of the source by 
\begin{equation}
    (V - I, I)_{\rm S,0} = (V - I, I)_{\rm S} - \Delta(V - I, I).
\end{equation}
Finally, using the color--surface brightness relation of \cite{Adams2018}, we obtain the angular source radius $\theta_*$. We summarize the measurements from the CMD analysis, the derived angular Einstein radius $\thetae$, and the lens--source relative proper motion $\mu_{\rm rel}$ in Table~\ref{source}. 
We note that the source of {\eventa} is $0.09$ magnitude redder than the red clump centroid and thus slightly off the sequence of evolved stars in the {\it HST} CMD. However, this offset is not significant compared to the dispersion in color at a similar magnitude in the {\it HST} stars. The source could well be a K4 type sub-giant in the bulge \citep{Bessell1988}.

\subsection{\rm Bayesian Analysis}\label{Bayesian}

Our Bayesian analysis applies the procedures and the Galactic model of \cite{OB191053}. The Galactic model is defined by the mass function of the lens, the stellar number density profile and the dynamical distributions. For the mass function of the lens, we choose the initial mass function (IMF) of \cite{Kroupa2001} with an upper limit of $1.3~M_{\odot}$ for disk lenses and $1.1~M_{\odot}$ for bulge lenses. For the stellar number density, we adopt the \cite{Zhu2017spitzer} model for bulge objects and the \cite{MB11262} model for disk objects. Regarding the kinematics, we adopt a rotation of $240~{\rm km~s}^{-1}$ \citep{Reid2014} and the velocity dispersion of \cite{KB180748} for disk lenses and the {\it Gaia} proper motion of red giant stars within $5'$ \citep{Gaia2016AA,Gaia2018AA} for bulge lenses as well as source stars. 

For each event, we create a sample of $10^8$ simulated events from the Galactic model and weight each simulated event, $i$, by
\begin{equation}
    \omega_{{\rm Gal},i} = \Gamma_{i} \mathcal{L}_{i}(\te^{\rm pri}) \mathcal{L}_{i}(\thetae^{\rm pri}), 
\end{equation}
where $\Gamma_{i}\varpropto\theta_{{\rm E},i}^{\rm pri} \times \mu_{{\rm rel},i}$ is the microlensing event rate, and $\mathcal{L}_{i}(\te^{\rm pri})$ and $\mathcal{L}_{i}(\thetae^{\rm pri})$ are the likelihoods of its inferred parameters given the distributions of these quantities, respectively. Here, $\te^{\rm pri}$ and $\thetae^{\rm pri}$ are the timescale and Einstein radius of the primary lens alone, respectively. They are a factor of $\sqrt{1+q}$ smaller than the values defined on the binary system.

Table \ref{phy} presents the inferred physical parameters of the lenses. For \eventa, the Bayesian analysis suggests a super-Earth-mass/sub-Neptune-mass planet about six times beyond the snow line of an ultracool dwarf near the M dwarf/brown-dwarf boundary (assuming a snow line radius $a_{\rm SL} = 2.7(M/M_{\odot})$~{\rm AU}, \citealt{snowline}). For \eventb\ and \eventc, the inferred companion masses exceed the mass limit of planets, with the former likely a low-mass star and the latter a brown dwarf.


\begin{table}
    \renewcommand\arraystretch{1.5}
    \centering
    \caption{CMD parameters, $\theta_*$, $\thetae$ and $\mu_{\rm rel}$ for the three events}
    \begin{tabular}{c c c c}
    \hline
    \hline
    Parameter & \eventa & \eventb & \eventc \\
    \hline
    $(V - I, I)_{\rm cl}$  & $(2.65 \pm 0.01, 16.24 \pm 0.03)$ & $(4.15 \pm 0.02, 17.73 \pm 0.03)$ & $(2.88 \pm 0.01, 16.49 \pm 0.04)$ \\
    $(V - I, I)_{\rm cl, 0}$ & $(1.06 \pm 0.03, 14.39 \pm 0.04)$ & $(1.06 \pm 0.03, 14.44 \pm 0.04)$ & $(1.06 \pm 0.03, 14.38 \pm 0.04)$  \\
    $(V - I, I)_{\rm S}$  & $(2.74 \pm 0.02, 18.370 \pm 0.023)$ & $(4.10 \pm 0.10, 18.778 \pm 0.009)$ & $(2.90 \pm 0.13, 16.94 \pm 0.07)$  \\
    $(V - I, I)_{\rm S,0}$  & $(1.15 \pm 0.04, 16.56 \pm 0.05)$ & $(1.01 \pm 0.11, 15.49 \pm 0.05)$ &  $(1.08 \pm 0.13, 14.83 \pm 0.08)$ \\
    $\theta_*$ ($\mu$as)   & $2.31 \pm 0.17$  & $3.64 \pm 0.65$  & $5.2 \pm 1.2$ \\
    $\thetae$ (mas) & $0.097 \pm 0.011$ &  $0.387 \pm 0.069$ & $>0.018$   \\ 
    $\mu_{\rm rel}$ (${\rm mas\,yr^{-1}}$) & $3.12 \pm 0.35$ & $4.66 \pm 0.84$ & $>0.61$ \\
    \hline
    \hline
    \end{tabular}
    \label{source}
\end{table}

\begin{table}
    \renewcommand\arraystretch{1.5}
    \centering
    \caption{Physical parameters of the lens systems, inferred from the Bayesian analysis. }
    \begin{tabular}{c c c c}
    \hline
    \hline
    Name & \eventa & \eventb & \eventc\\
    \hline
    $M_{\rm 1}[M_{\odot}]$ & $0.10_{-0.05}^{+0.13}$ & $0.41_{-0.23}^{+0.19}$ & $0.23_{-0.14}^{+0.28}$ \\
    
    $M_{\rm 2}$ & $6.4_{-2.8}^{+5.5}~M_{\oplus}$ & $0.24_{-0.14}^{+0.12}~M_{\odot}$ & $23.1_{-14.4}^{+30.2}~M_J$ \\
    
    $D_{\rm L}$[kpc] & $7.7_{-0.6}^{+0.6}$ & $6.9_{-1.5}^{+0.7}$ & $7.2_{-1.4}^{+0.7}$ \\
    
    $a_{\perp}$[au] & $1.8_{-0.2}^{+0.2}$ & $1.5_{-0.3}^{+0.3}$ & $0.6_{-0.2}^{+0.2}$ \\
    
    $\mu_{\rm rel}$[${\rm mas\,yr^{-1}}$] & $3.2_{-0.3}^{+0.3}$ & $4.7_{-0.8}^{+0.8}$ & $7.2_{-2.6}^{+3.4}$ \\
    \hline
    \hline
    \end{tabular}\\
    \label{phy}
\end{table}
\section{Discussion}\label{sec:discussion}

\begin{figure}
    \centering
    \includegraphics[width=0.70\columnwidth]{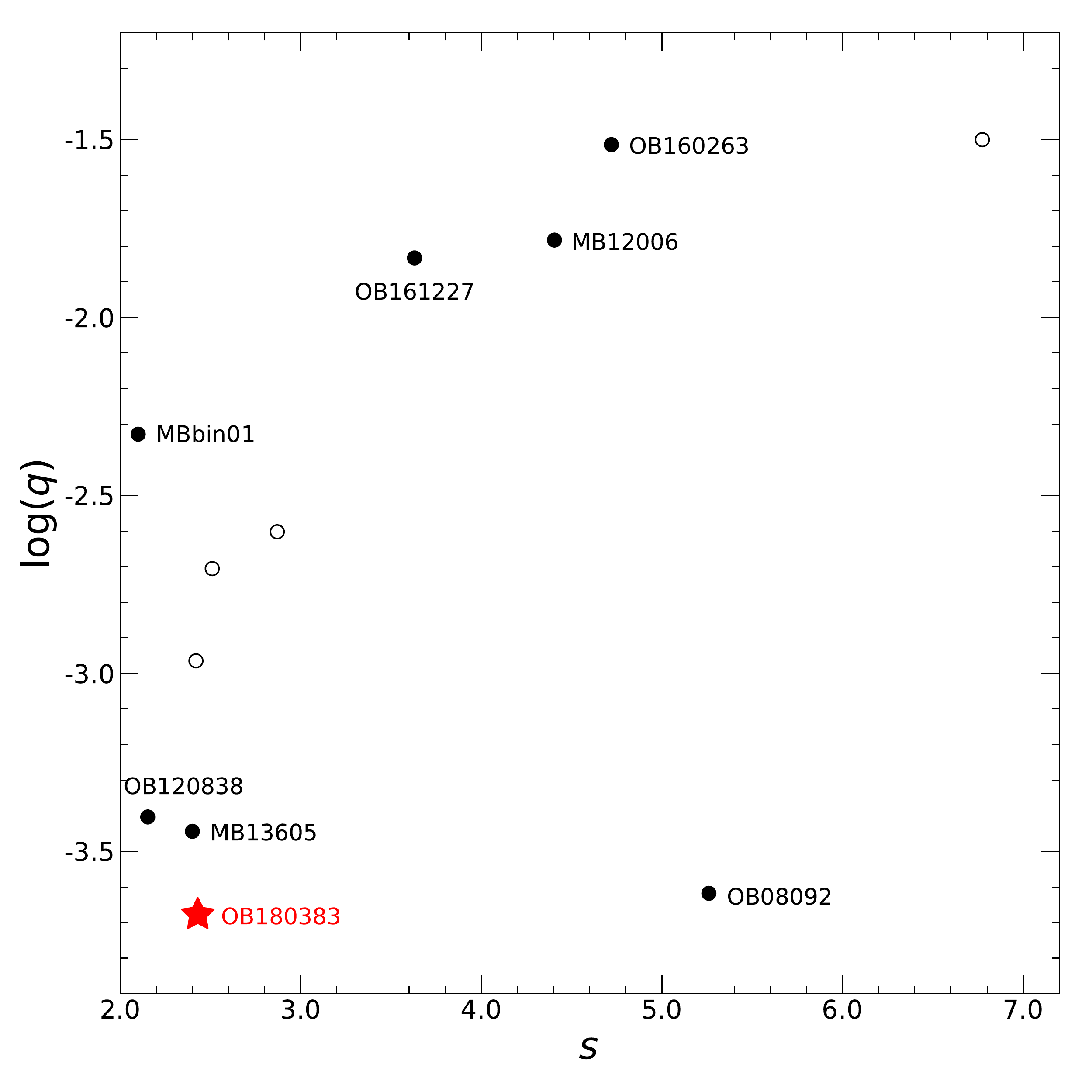}
    \caption{ All known microlensing planets with $s>2.0$. The red asterisk marks \eventa\ from the present work. The solid dots are events with ``unique'' solutions (i.e., no degenerate solution within $\Delta\chi^2<10$) and the open circles are events with degenerate solutions. The abridged event name is shown next to those with unique solutions.}
    \label{sq}
\end{figure}

In this work, we have presented the discovery and characterization of three microlensing systems that were originally identified to contain candidates for wide-orbit ($s>2$) planets. Detailed modeling has revealed that the lens system in \eventa\ indeed contains a wide-orbit planet with projected separation $s=2.45$. With a planet-to-star mass ratio $q=2.1\times10^{-4}$, it is also the wide-orbit planet with so far the lowest mass ratio (see Figure~\ref{sq}). The other two events, \eventb\ and\ \eventc, are shown to be produced by close ($s=0.55$ and $0.41$) binaries with relatively large mass ratios ($q=0.6$ and $0.1$). This highlights the importance of detailed light curve modeling in identifying (close- and wide-orbit) microlensing planets.

The wide-orbit planets found by microlensing are shown in Figure~\ref{sq}.
\footnote{Our sample differs from that of \citet{Poleski:2021} by the exclusion of event OGLE-2011-BLG-0173, for which the binary source model could not be ruled out by $\Delta\chi^2>10$ \citep{OB110173}.}
These planets were mostly detected via planetary anomalies that were well separated from the primary lensing signals of the host stars (e.g., Figure~\ref{lc_0383}), although the wide-orbit nature of the planets could also be revealed in the careful investigation of short-timescale binary events (e.g., MOA-bin-1 and OGLE-2016-BLG-1227, \citealt{Bennett:2012,OB161227}). Events with these characteristics are rarely targets of follow-up observations, and thus the discovery of wide-orbit planets relies almost entirely on microlensing survey observations. Out of the eight known wide-orbit planets shown in Figure~\ref{sq}, five (OGLE-2008-BLG-092, MOA-2012-BLG-006, OGLE-2012-BLG-0838, MOA-2013-BLG-605, and OGLE-2016-BLG-0263) are included in the sample of \citet{Poleski:2021}, one (MOA-bin-1) was only detected in MOA data, and the remaining two (OGLE-2016-BLG-1227 and OGLE-2018-BLG-0383) could not have been detected without the KMTNet data. Because the anomalous feature is either small or well separated from the main peak, the majority of the wide-orbit planets could only be detected via systematic searches for anomalous events. Now with the successful implementation of systematic anomaly search in the KMTNet data \citep{OB191053}, we expect that the sample of wide-orbit planets will expand more rapidly.

It is also worth noting that the source stars of microlensing events containing wide-orbit planets are all evolved stars. These stars are relatively bright and have relatively large size. The former ensures better photometric precision and thus the detection for more subtle deviations, whereas the latter leads to a prolonged duration of the anomalous feature. Future systematic search and statistical studies of wide-orbit planets may target events with evolved stars. This so-called ``Hollywood'' strategy of ``following the big stars'', was originally advocated by \citet{Hollywood}.

\eventb\ reveals some interesting characteristics that are worth reporting, even though it is not of planetary nature.
Unlike the majority of anomalous events found by AnomalyFinder \citep{OB191053}, the anomalous feature in \eventb\ was first recognized by the KMTNet EventFinder algorithm \citep{Kim2018a} as a short-timescale event, and the lensing signal from the primary star was later identified by the AnomalyFinder algorithm as the ``anomaly.'' This is because the anomalous feature, even though with a shorter duration, has a much larger amplitude than the lensing signal from the primary star. 
Such a feature is also seen in events with wide-orbit planets \citep{Bennett:2012,OB161227}. In the extreme case of OGLE-2016-BLG-1227 \citep{OB161227}, the light curve appears to be a short-lived 1L1S event affected by severe finite-source effect, and there is no obvious signal from the host star. Only with a detailed analysis was the presence of a distant host revealed from the $\sim$0.03\,mag perturbation to the 1L1S model \citep{OB161227}. Such events again highlight the importance of dense and continuous coverage of observations and detailed light curve modeling in studies of wide-orbit planets.




\section*{Acknowledgements}
We acknowledge the science research grants from the China Manned Space Project with No.\ CMS-CSST-2021-A11.
W.Zang, H.Y., S.M. and X.Z. acknowledge support by the National Science Foundation of China (Grant No.\ 11821303 and 11761131004). 
This research has made use of the KMTNet system operated by the Korea Astronomy and Space Science Institute (KASI) and the data were obtained at three host sites of CTIO in Chile, SAAO in South Africa, and SSO in Australia. 
Work by JCY was supported by JPL grant 1571564.
Work by C.H. was supported by the grants of National Research Foundation of Korea (2019R1A2C2085965 and 2020R1A4A2002885).
Work by RP was supported by Polish National Agency for Academic Exchange grant ``Polish Returns 2019.'' 
This research has made use of the NASA Exoplanet Archive, which is operated by the California Institute of Technology, under contract with the National Aeronautics and Space Administration under the Exoplanet Exploration Program.

\section*{Data Availability}

Data used in the light curve analysis will be provided along with publication.





\bibliographystyle{mnras}
\bibliography{Zang.bib} 

\begin{thebibliography}{}
\makeatletter
\relax
\def\mn@urlcharsother{\let\do\@makeother \do\$\do\&\do\#\do\^\do\_\do\%\do\~}
\def\mn@doi{\begingroup\mn@urlcharsother \@ifnextchar [ {\mn@doi@}
  {\mn@doi@[]}}
\def\mn@doi@[#1]#2{\def\@tempa{#1}\ifx\@tempa\@empty \href
  {http://dx.doi.org/#2} {doi:#2}\else \href {http://dx.doi.org/#2} {#1}\fi
  \endgroup}
\def\mn@eprint#1#2{\mn@eprint@#1:#2::\@nil}
\def\mn@eprint@arXiv#1{\href {http://arxiv.org/abs/#1} {{\tt arXiv:#1}}}
\def\mn@eprint@dblp#1{\href {http://dblp.uni-trier.de/rec/bibtex/#1.xml}
  {dblp:#1}}
\def\mn@eprint@#1:#2:#3:#4\@nil{\def\@tempa {#1}\def\@tempb {#2}\def\@tempc
  {#3}\ifx \@tempc \@empty \let \@tempc \@tempb \let \@tempb \@tempa \fi \ifx
  \@tempb \@empty \def\@tempb {arXiv}\fi \@ifundefined
  {mn@eprint@\@tempb}{\@tempb:\@tempc}{\expandafter \expandafter \csname
  mn@eprint@\@tempb\endcsname \expandafter{\@tempc}}}

\bibitem[\protect\citeauthoryear{{Adams}, {Boyajian}  \& {von Braun}}{{Adams}
  et~al.}{2018}]{Adams2018}
{Adams} A.~D.,  {Boyajian} T.~S.,   {von Braun} K.,  2018, \mn@doi [\mnras]
  {10.1093/mnras/stx2367}, \href
  {http://adsabs.harvard.edu/abs/2018MNRAS.473.3608A} {473, 3608}

\bibitem[\protect\citeauthoryear{{Alard} \& {Lupton}}{{Alard} \&
  {Lupton}}{1998}]{Alard1998}
{Alard} C.,  {Lupton} R.~H.,  1998, \mn@doi [\apj] {10.1086/305984}, \href
  {http://adsabs.harvard.edu/abs/1998ApJ...503..325A} {503, 325}

\bibitem[\protect\citeauthoryear{{Albrow} et~al.,}{{Albrow}
  et~al.}{2009}]{pysis}
{Albrow} M.~D.,  et~al., 2009, \mn@doi [\mnras]
  {10.1111/j.1365-2966.2009.15098.x}, \href
  {http://adsabs.harvard.edu/abs/2009MNRAS.397.2099A} {397, 2099}

\bibitem[\protect\citeauthoryear{{An} et~al.,}{{An} et~al.}{2002}]{An2002}
{An} J.~H.,  et~al., 2002, \mn@doi [\apj] {10.1086/340191}, \href
  {http://adsabs.harvard.edu/abs/2002ApJ...572..521A} {572, 521}

\bibitem[\protect\citeauthoryear{{Batista} et~al.,}{{Batista}
  et~al.}{2011}]{MB09387}
{Batista} V.,  et~al., 2011, \mn@doi [\aap] {10.1051/0004-6361/201016111},
  \href {http://adsabs.harvard.edu/abs/2011A%26A...529A.102B} {529, A102}

\bibitem[\protect\citeauthoryear{{Bennett} et~al.,}{{Bennett}
  et~al.}{2008}]{MB07192}
{Bennett} D.~P.,  et~al., 2008, \mn@doi [\apj] {10.1086/589940}, \href
  {http://adsabs.harvard.edu/abs/2008ApJ...684..663B} {684, 663}

\bibitem[\protect\citeauthoryear{{Bennett} et~al.,}{{Bennett}
  et~al.}{2012}]{Bennett:2012}
{Bennett} D.~P.,  et~al., 2012, \mn@doi [\apj] {10.1088/0004-637X/757/2/119},
  \href {https://ui.adsabs.harvard.edu/abs/2012ApJ...757..119B} {757, 119}

\bibitem[\protect\citeauthoryear{{Bennett} et~al.,}{{Bennett}
  et~al.}{2014}]{MB11262}
{Bennett} D.~P.,  et~al., 2014, \mn@doi [\apj] {10.1088/0004-637X/785/2/155},
  \href {https://ui.adsabs.harvard.edu/abs/2014ApJ...785..155B} {785, 155}

\bibitem[\protect\citeauthoryear{{Bensby} et~al.,}{{Bensby}
  et~al.}{2013}]{Bensby2013}
{Bensby} T.,  et~al., 2013, \mn@doi [\aap] {10.1051/0004-6361/201220678}, \href
  {http://adsabs.harvard.edu/abs/2013A%26A...549A.147B} {549, A147}

\bibitem[\protect\citeauthoryear{{Bessell} \& {Brett}}{{Bessell} \&
  {Brett}}{1988}]{Bessell1988}
{Bessell} M.~S.,  {Brett} J.~M.,  1988, \mn@doi [\pasp] {10.1086/132281}, \href
  {http://adsabs.harvard.edu/abs/1988PASP..100.1134B} {100, 1134}

\bibitem[\protect\citeauthoryear{{Bozza}}{{Bozza}}{2010}]{Bozza2010}
{Bozza} V.,  2010, \mn@doi [\mnras] {10.1111/j.1365-2966.2010.17265.x}, \href
  {http://adsabs.harvard.edu/abs/2010MNRAS.408.2188B} {408, 2188}

\bibitem[\protect\citeauthoryear{{Bozza}, {Bachelet}, {Bartoli{\'c}}, {Heintz},
  {Hoag}  \& {Hundertmark}}{{Bozza} et~al.}{2018}]{Bozza2018}
{Bozza} V.,  {Bachelet} E.,  {Bartoli{\'c}} F.,  {Heintz} T.~M.,  {Hoag} A.~R.,
    {Hundertmark} M.,  2018, \mn@doi [\mnras] {10.1093/mnras/sty1791}, \href
  {https://ui.adsabs.harvard.edu/abs/2018MNRAS.479.5157B} {479, 5157}

\bibitem[\protect\citeauthoryear{{Dominik}}{{Dominik}}{1999}]{Dominik1999}
{Dominik} M.,  1999, \aap, \href
  {http://adsabs.harvard.edu/abs/1999A%26A...349..108D} {349, 108}

\bibitem[\protect\citeauthoryear{{Dong} et~al.,}{{Dong}
  et~al.}{2009}]{OB050071D}
{Dong} S.,  et~al., 2009, \mn@doi [\apj] {10.1088/0004-637X/695/2/970}, \href
  {http://adsabs.harvard.edu/abs/2009ApJ...695..970D} {695, 970}

\bibitem[\protect\citeauthoryear{{Foreman-Mackey}, {Hogg}, {Lang}  \&
  {Goodman}}{{Foreman-Mackey} et~al.}{2013}]{emcee}
{Foreman-Mackey} D.,  {Hogg} D.~W.,  {Lang} D.,   {Goodman} J.,  2013, \mn@doi
  [\pasp] {10.1086/670067}, \href
  {http://adsabs.harvard.edu/abs/2013PASP..125..306F} {125, 306}

\bibitem[\protect\citeauthoryear{{Gaia Collaboration} et~al.,}{{Gaia
  Collaboration} et~al.}{2016}]{Gaia2016AA}
{Gaia Collaboration} et~al., 2016, \mn@doi [\aap]
  {10.1051/0004-6361/201629272}, \href
  {http://adsabs.harvard.edu/abs/2016A%26A...595A...1G} {595, A1}

\bibitem[\protect\citeauthoryear{{Gaia Collaboration} et~al.,}{{Gaia
  Collaboration} et~al.}{2018}]{Gaia2018AA}
{Gaia Collaboration} et~al., 2018, \mn@doi [\aap]
  {10.1051/0004-6361/201833051}, \href
  {http://adsabs.harvard.edu/abs/2018A%26A...616A...1G} {616, A1}

\bibitem[\protect\citeauthoryear{{Gaudi}}{{Gaudi}}{1998}]{Gaudi1998}
{Gaudi} B.~S.,  1998, \mn@doi [\apj] {10.1086/306256}, \href
  {http://adsabs.harvard.edu/abs/1998ApJ...506..533G} {506, 533}

\bibitem[\protect\citeauthoryear{{Gaudi} \& {Gould}}{{Gaudi} \&
  {Gould}}{1997}]{GaudiGould1997}
{Gaudi} B.~S.,  {Gould} A.,  1997, \mn@doi [\apj] {10.1086/304491}, \href
  {https://ui.adsabs.harvard.edu/abs/1997ApJ...486...85G} {486, 85}

\bibitem[\protect\citeauthoryear{{Gould}}{{Gould}}{1992}]{Gould1992}
{Gould} A.,  1992, \mn@doi [\apj] {10.1086/171443}, \href
  {http://adsabs.harvard.edu/abs/1992ApJ...392..442G} {392, 442}

\bibitem[\protect\citeauthoryear{{Gould}}{{Gould}}{1994}]{1994ApJ...421L..75G}
{Gould} A.,  1994, \mn@doi [\apjl] {10.1086/187191}, \href
  {http://adsabs.harvard.edu/abs/1994ApJ...421L..75G} {421, L75}

\bibitem[\protect\citeauthoryear{{Gould}}{{Gould}}{1997}]{Hollywood}
{Gould} A.,  1997, in {Ferlet} R.,  {Maillard} J.-P.,   {Raban} B.,  eds,
  Variables Stars and the Astrophysical Returns of the Microlensing Surveys.
  p.~125 (\mn@eprint {arXiv} {astro-ph/9608045})

\bibitem[\protect\citeauthoryear{{Gould}}{{Gould}}{2000}]{Gould2000}
{Gould} A.,  2000, \mn@doi [\apj] {10.1086/317037}, \href
  {http://adsabs.harvard.edu/abs/2000ApJ...542..785G} {542, 785}

\bibitem[\protect\citeauthoryear{{Gould}}{{Gould}}{2004}]{Gouldpies2004}
{Gould} A.,  2004, \mn@doi [\apj] {10.1086/382782}, \href
  {http://adsabs.harvard.edu/abs/2004ApJ...606..319G} {606, 319}

\bibitem[\protect\citeauthoryear{{Gould} \& {Gaucherel}}{{Gould} \&
  {Gaucherel}}{1997}]{Gould1997}
{Gould} A.,  {Gaucherel} C.,  1997, \mn@doi [\apj] {10.1086/303751}, \href
  {https://ui.adsabs.harvard.edu/abs/1997ApJ...477..580G} {477, 580}

\bibitem[\protect\citeauthoryear{{Gould} \& {Loeb}}{{Gould} \&
  {Loeb}}{1992}]{Andy1992}
{Gould} A.,  {Loeb} A.,  1992, \mn@doi [\apj] {10.1086/171700}, \href
  {http://adsabs.harvard.edu/abs/1992ApJ...396..104G} {396, 104}

\bibitem[\protect\citeauthoryear{{Griest} \& {Safizadeh}}{{Griest} \&
  {Safizadeh}}{1998}]{Griest1998}
{Griest} K.,  {Safizadeh} N.,  1998, \mn@doi [\apj] {10.1086/305729}, \href
  {http://adsabs.harvard.edu/abs/1998ApJ...500...37G} {500, 37}

\bibitem[\protect\citeauthoryear{{Han} et~al.,}{{Han} et~al.}{2020a}]{OB161227}
{Han} C.,  et~al., 2020a, \mn@doi [\aj] {10.3847/1538-3881/ab6a9f}, \href
  {https://ui.adsabs.harvard.edu/abs/2020AJ....159...91H} {159, 91}

\bibitem[\protect\citeauthoryear{{Han} et~al.,}{{Han} et~al.}{2020b}]{KB180748}
{Han} C.,  et~al., 2020b, \mn@doi [\aap] {10.1051/0004-6361/202038173}, \href
  {https://ui.adsabs.harvard.edu/abs/2020A&A...641A.105H} {641, A105}

\bibitem[\protect\citeauthoryear{{Holtzman}, {Watson}, {Baum}, {Grillmair},
  {Groth}, {Light}, {Lynds}  \& {O'Neil}}{{Holtzman} et~al.}{1998}]{HSTCMD}
{Holtzman} J.~A.,  {Watson} A.~M.,  {Baum} W.~A.,  {Grillmair} C.~J.,  {Groth}
  E.~J.,  {Light} R.~M.,  {Lynds} R.,   {O'Neil} Jr. E.~J.,  1998, \mn@doi
  [\aj] {10.1086/300336}, \href
  {http://adsabs.harvard.edu/abs/1998AJ....115.1946H} {115, 1946}

\bibitem[\protect\citeauthoryear{{Hwang} et~al.,}{{Hwang}
  et~al.}{2021}]{KB190253}
{Hwang} K.-H.,  et~al., 2021, arXiv e-prints, \href
  {https://ui.adsabs.harvard.edu/abs/2021arXiv210606686H} {p. arXiv:2106.06686}

\bibitem[\protect\citeauthoryear{{Kennedy} \& {Kenyon}}{{Kennedy} \&
  {Kenyon}}{2008}]{snowline}
{Kennedy} G.~M.,  {Kenyon} S.~J.,  2008, \mn@doi [\apj] {10.1086/524130}, \href
  {http://adsabs.harvard.edu/abs/2008ApJ...673..502K} {673, 502}

\bibitem[\protect\citeauthoryear{{Kim} et~al.,}{{Kim} et~al.}{2016}]{KMT2016}
{Kim} S.-L.,  et~al., 2016, \mn@doi [Journal of Korean Astronomical Society]
  {10.5303/JKAS.2016.49.1.037}, \href
  {http://adsabs.harvard.edu/abs/2016JKAS...49...37K} {49, 37}

\bibitem[\protect\citeauthoryear{{Kim} et~al.,}{{Kim} et~al.}{2018}]{Kim2018a}
{Kim} D.-J.,  et~al., 2018, \mn@doi [\aj] {10.3847/1538-3881/aaa47b}, \href
  {http://adsabs.harvard.edu/abs/2018AJ....155...76K} {155, 76}

\bibitem[\protect\citeauthoryear{{Kroupa}}{{Kroupa}}{2001}]{Kroupa2001}
{Kroupa} P.,  2001, \mn@doi [\mnras] {10.1046/j.1365-8711.2001.04022.x}, \href
  {http://adsabs.harvard.edu/abs/2001MNRAS.322..231K} {322, 231}

\bibitem[\protect\citeauthoryear{{Mao} \& {Paczynski}}{{Mao} \&
  {Paczynski}}{1991}]{Shude1991}
{Mao} S.,  {Paczynski} B.,  1991, \mn@doi [\apjl] {10.1086/186066}, \href
  {http://adsabs.harvard.edu/abs/1991ApJ...374L..37M} {374, L37}

\bibitem[\protect\citeauthoryear{{Mayor} \& {Queloz}}{{Mayor} \&
  {Queloz}}{1995}]{Mayor:1995}
{Mayor} M.,  {Queloz} D.,  1995, \mn@doi [\nat] {10.1038/378355a0}, \href
  {https://ui.adsabs.harvard.edu/abs/1995Natur.378..355M} {378, 355}

\bibitem[\protect\citeauthoryear{{Nataf} et~al.,}{{Nataf}
  et~al.}{2013}]{Nataf2013}
{Nataf} D.~M.,  et~al., 2013, \mn@doi [\apj] {10.1088/0004-637X/769/2/88},
  \href {http://adsabs.harvard.edu/abs/2013ApJ...769...88N} {769, 88}

\bibitem[\protect\citeauthoryear{{Nataf} et~al.,}{{Nataf}
  et~al.}{2016}]{Nataf2016}
{Nataf} D.~M.,  et~al., 2016, \mn@doi [\mnras] {10.1093/mnras/stv2843}, \href
  {http://adsabs.harvard.edu/abs/2016MNRAS.456.2692N} {456, 2692}

\bibitem[\protect\citeauthoryear{{Nemiroff} \& {Wickramasinghe}}{{Nemiroff} \&
  {Wickramasinghe}}{1994}]{Nemiroff1994}
{Nemiroff} R.~J.,  {Wickramasinghe} W.~A.~D.~T.,  1994, \mn@doi [\apjl]
  {10.1086/187265}, \href {http://adsabs.harvard.edu/abs/1994ApJ...424L..21N}
  {424, L21}

\bibitem[\protect\citeauthoryear{{Paczy{\'n}ski}}{{Paczy{\'n}ski}}{1986}]{Paczynski1986}
{Paczy{\'n}ski} B.,  1986, \mn@doi [\apj] {10.1086/164140}, \href
  {http://adsabs.harvard.edu/abs/1986ApJ...304....1P} {304, 1}

\bibitem[\protect\citeauthoryear{{Poleski} et~al.,}{{Poleski}
  et~al.}{2014}]{OB08092}
{Poleski} R.,  et~al., 2014, \mn@doi [\apj] {10.1088/0004-637X/795/1/42}, \href
  {https://ui.adsabs.harvard.edu/abs/2014ApJ...795...42P} {795, 42}

\bibitem[\protect\citeauthoryear{{Poleski} et~al.,}{{Poleski}
  et~al.}{2018}]{OB110173}
{Poleski} R.,  et~al., 2018, \mn@doi [\aj] {10.3847/1538-3881/aad45e}, \href
  {https://ui.adsabs.harvard.edu/abs/2018AJ....156..104P} {156, 104}

\bibitem[\protect\citeauthoryear{{Poleski} et~al.,}{{Poleski}
  et~al.}{2021}]{Poleski:2021}
{Poleski} R.,  et~al., 2021, \mn@doi [\actaa] {10.32023/0001-5237/71.1.1},
  \href {https://ui.adsabs.harvard.edu/abs/2021AcA....71....1P} {71, 1}

\bibitem[\protect\citeauthoryear{{Reid} et~al.,}{{Reid}
  et~al.}{2014}]{Reid2014}
{Reid} M.~J.,  et~al., 2014, \mn@doi [\apj] {10.1088/0004-637X/783/2/130},
  \href {https://ui.adsabs.harvard.edu/abs/2014ApJ...783..130R} {783, 130}

\bibitem[\protect\citeauthoryear{{Skowron} et~al.,}{{Skowron}
  et~al.}{2011}]{OB09020}
{Skowron} J.,  et~al., 2011, \mn@doi [\apj] {10.1088/0004-637X/738/1/87}, \href
  {http://adsabs.harvard.edu/abs/2011ApJ...738...87S} {738, 87}

\bibitem[\protect\citeauthoryear{{Tomaney} \& {Crotts}}{{Tomaney} \&
  {Crotts}}{1996}]{Tomaney1996}
{Tomaney} A.~B.,  {Crotts} A. P.~S.,  1996, \mn@doi [\aj] {10.1086/118228},
  \href {https://ui.adsabs.harvard.edu/abs/1996AJ....112.2872T} {112, 2872}

\bibitem[\protect\citeauthoryear{{Udalski}}{{Udalski}}{2003}]{Udalski2003}
{Udalski} A.,  2003, \actaa, \href
  {http://adsabs.harvard.edu/abs/2003AcA....53..291U} {53, 291}

\bibitem[\protect\citeauthoryear{{Udalski}, {Szymanski}, {Kaluzny}, {Kubiak},
  {Mateo}, {Krzeminski}  \& {Paczynski}}{{Udalski} et~al.}{1994}]{Udalski1994}
{Udalski} A.,  {Szymanski} M.,  {Kaluzny} J.,  {Kubiak} M.,  {Mateo} M.,
  {Krzeminski} W.,   {Paczynski} B.,  1994, \actaa, \href
  {http://adsabs.harvard.edu/abs/1994AcA....44..227U} {44, 227}

\bibitem[\protect\citeauthoryear{{Udalski}, {Szyma{\'n}ski}  \&
  {Szyma{\'n}ski}}{{Udalski} et~al.}{2015}]{OGLEIV}
{Udalski} A.,  {Szyma{\'n}ski} M.~K.,   {Szyma{\'n}ski} G.,  2015, \actaa,
  \href {http://adsabs.harvard.edu/abs/2015AcA....65....1U} {65, 1}

\bibitem[\protect\citeauthoryear{{Winn} \& {Fabrycky}}{{Winn} \&
  {Fabrycky}}{2015}]{WinnFabrycky:2015}
{Winn} J.~N.,  {Fabrycky} D.~C.,  2015, \mn@doi [\araa]
  {10.1146/annurev-astro-082214-122246}, \href
  {https://ui.adsabs.harvard.edu/abs/2015ARA&A..53..409W} {53, 409}

\bibitem[\protect\citeauthoryear{{Witt} \& {Mao}}{{Witt} \&
  {Mao}}{1994}]{Shude1994}
{Witt} H.~J.,  {Mao} S.,  1994, \mn@doi [\apj] {10.1086/174426}, \href
  {http://adsabs.harvard.edu/abs/1994ApJ...430..505W} {430, 505}

\bibitem[\protect\citeauthoryear{{Wozniak}}{{Wozniak}}{2000}]{Wozniak2000}
{Wozniak} P.~R.,  2000, \actaa, \href
  {http://adsabs.harvard.edu/abs/2000AcA....50..421W} {50, 421}

\bibitem[\protect\citeauthoryear{{Yoo} et~al.,}{{Yoo} et~al.}{2004}]{Yoo2004}
{Yoo} J.,  et~al., 2004, \mn@doi [\apj] {10.1086/381241}, \href
  {http://adsabs.harvard.edu/abs/2004ApJ...603..139Y} {603, 139}

\bibitem[\protect\citeauthoryear{{Zang} et~al.,}{{Zang}
  et~al.}{2021}]{OB191053}
{Zang} W.,  et~al., 2021, \mn@doi [\aj] {10.3847/1538-3881/ac12d4}, \href
  {https://ui.adsabs.harvard.edu/abs/2021AJ....162..163Z} {162, 163}

\bibitem[\protect\citeauthoryear{{Zhu} \& {Dong}}{{Zhu} \&
  {Dong}}{2021}]{ZhuDong:2021}
{Zhu} W.,  {Dong} S.,  2021, arXiv e-prints, \href
  {https://ui.adsabs.harvard.edu/abs/2021arXiv210302127Z} {p. arXiv:2103.02127}

\bibitem[\protect\citeauthoryear{{Zhu} et~al.,}{{Zhu}
  et~al.}{2017}]{Zhu2017spitzer}
{Zhu} W.,  et~al., 2017, \mn@doi [\aj] {10.3847/1538-3881/aa8ef1}, \href
  {http://adsabs.harvard.edu/abs/2017AJ....154..210Z} {154, 210}

\makeatother
\end{thebibliography}







\end{CJK*}

\bsp	
\label{lastpage}
\end{document}